\documentstyle[12pt,epsf]{article}
\renewcommand{\thefootnote}{\fnsymbol{footnote}}
\newcommand{\be}{\begin{equation}}
\newcommand{\ee}{\end{equation}}
\newcommand{\bear}{\begin{eqnarray}}
\newcommand{\ear}{\end{eqnarray}}

\newcommand{\cL}{\cal L}
\newcommand{\cS}{\cal S}
\newcommand{\rw}{\rm w}
\newcommand{\rc}{\rm c}
\newcommand\eins{  1\!{\rm l}  }
\newcommand\DF{\bf F}

\newsavebox{\LSIM}
\sbox{\LSIM}{\raisebox{-1ex}{$\ \stackrel{\textstyle<}{\sim}\ $}}
\newcommand{\lsim}{\usebox{\LSIM}}
\newsavebox{\GSIM}
\sbox{\GSIM}{\raisebox{-1ex}{$\ \stackrel{\textstyle>}{\sim}\ $}}
\newcommand{\gsim}{\usebox{\GSIM}}
\renewcommand{\theequation}{\arabic{section}.\arabic{equation}}
\begin{document}
\begin{titlepage}
\begin{flushright}
HD-THEP-98-16\\
MZ-TH/98-15\\
hep-ph/9804441
\end{flushright}
\vspace{.5cm}
\begin{center}
{\bf\LARGE Nonperturbative Contributions}\\
\vspace{.3cm}
{\bf\LARGE to the Hot Electroweak Potential}\\
\vspace{.3cm}
{\bf\LARGE and the Crossover}\\
\vspace{.5cm}
Stephan J. Huber$^1$\footnote{e-mail: s.huber@thphys.uni-heidelberg.de}\\
Andreas Laser$^2$\footnote{e-mail:
andreas.laser@arthurandersen.com}\\
Martin Reuter$^3$\footnote{e-mail:
reuter@thep.physik.uni-mainz.de}\\
Michael G. Schmidt$^4$\footnote{e-mail: m.g.
schmidt@thphys.uni-heidelberg.de}\\
\end{center}
\bigskip\noindent
\noindent
$^{1,4}$ Institut f\"ur Theoretische Physik der Universit\"at
Heidelberg,

Philosophenweg 16, D-69120 Heidelberg
\\
$^2$ Arthur Andersen GmbH, Mergenthalerallee 10-12, D-65760 Eschborn\\
\noindent
$^3$ Institut f\"ur  Physik,  Universit\"at
Mainz, Staudingerweg 7, D-55099 Mainz
\\

\vspace{.5cm}
\begin{abstract}
We discuss nonperturbative contributions to the 3-dimensional
one-loop effective potential of the electroweak theory at high
temperatures in the framework of the stochastic vacuum model.
It assumes a gauge-field background with Gaussian correlations
which leads to confinement. The instability of $<F^2>=0$
in Yang-Mills theory appears for small Higgs expectation value 
$<\phi^2>$ in an IR
regularized form. The gauge boson propagator obtains a positive
momentum-dependent ``diamagnetic'' effective (mass)$^2$ due to
confinement effects and a negative one due to ``paramagnetic'' spin-spin
interactions which are related to the $<F^2>=0$ instability.
Numerical evaluation of an approximate effective
potential containing these masses shows qualitatively the
fading away of the first-order phase transition with increasing Higgs mass 
which was observed in
lattice calculations. The crossover point can be roughly determined
postulating that the effective $\phi^4$ and $\phi^2$ terms vanish there.
\end{abstract}
\end{titlepage}
\renewcommand{\thefootnote}{\arabic{footnote}}
\setcounter{footnote}{0}
\section{Introduction}
The phase diagram of the electroweak standard model at
high temperatures $T$ of about 100 GeV can be well described by the
3-dimensional effective Higgs field gauge theory which is obtained
perturbatively by ``integrating out'' heavy degrees of
freedom, i.e. by matching a set of amplitudes in 4 and 3 dimensions
\cite{1,2}:
\be\label{1.1}
{\cL}_3=\frac{1}{4}F^a_{\mu\nu}F^a_{\mu\nu}
+(D_\mu\phi)^+(D_\mu\phi)+m_3^2\phi^+\phi+
\lambda_3(\phi^+\phi)^2.\ee
Here higher derivative terms are neglected. This is adequate to
order  $g^3_{\rw}$, where $g_{\rw}$ is the weak coupling constant
in $d=4$. The SU(2) Yang-Mills action $\frac{1}{4}F^2$
and the covariant derivative $D_\mu$ contain a gauge coupling $g^2_3$
proportional to $g^2_{\rw}T$. The U(1) part of the standard
theory is neglected here
since it is not important for our considerations, but it could be
easily included. Furthermore, $m_3$ and $\lambda_3$ are the effective  
$T$-dependent mass and coupling constant of the dimensionally
reduced theory, respectively.

The dimensionless ratio $g^2_3/m_{\rm IR}$ with some typical
physical infrared (\rm IR) scale
$m_{\rm IR}$ is not small in general. Indeed, the gauge coupling
in a Wilson-type effective action increases strongly
with the IR cut-off going to zero \cite{2a}.
Thus the further use of the
perturbation theory based upon the Lagrangian (\ref{1.1}) is very
dangerous (like in QCD) and it is only consistent if the scale
$<\phi^+\phi>$ is not too small. Besides this, one
also has to check if the quantity which one tries to calculate
perturbatively
is not dominated by nonperturbative
contributions, in particular if the former turns out to be small, as
in the case of the effective potential to be discussed.

Since the fermions are integrated out already,
the action (\ref{1.1}) can be easily  put on a lattice
\cite{3,4}.
The critical temperature $T_{\rc}$ of the phase transition,
${v(T_{\rc})}/{T_{\rc}}$ with $v(T_{\rc})$ the Higgs field minimum, the
latent heat, and the surface tension can be calculated (``measured'')
with good accuracy. The dimensionless quantities
\be\label{1.2}
y=\frac{m^2_3}{g^4_3},\quad x=\frac{\lambda_3}{g^2_3}\ee
span the phase diagram; $y$ is connected to $(T-T_{\rc})$
and $x$ describes the strength of the phase transition and
depends on the Higgs mass. For
$x\lsim0.03-0.04$ the first order phase transition obtained in two-loop
perturbation theory compares very well with lattice calculations
\cite{3} of the above-mentioned quantities.

For larger $x$ values there
are deviations which become stronger for quantities of decreasing
perturbative order in $g_{\rw}$.
In the above list, the surface tension shows
the worst failure of a perturbative calculation \cite{5}.
For $x\lsim0.11$ (corresponding to
$m_H\sim m_W$) lattice calculations show \cite{3} that
the phase transition fades away, there is a crossover. On
the other hand, a two-loop perturbative treatment \cite{1,6}
gives a first-order phase transition.
Perturbation theory can be formally applied since
$g^2_3/g_{\rw}v(T)$ is still rather small in this $x$-range;
but the effective potential
becomes very shallow and $v(T_{\rc})/T_{\rc}$ is quite small,
so that the phase transition
is very weakly first order. We conclude that in such a situation
nonperturbative contributions to the effective potential are
essential.

In this paper we develop a physical picture for such nonperturbative
contributions. We will present analytic indications for an instability
of the vacuum of the 3-dimensional theory (\ref{1.1}) at $<F^2>=0$
in case of small Higgs vevs. We then will develop a up to now 
qualitative model for the $<F^2>\neq 0$ vacuum contribution
to the electroweak potential. This will lead to an understanding of
the crossover.

Lattice studies reveal that in the high temperature phase of the
effective theory (\ref{1.1}), i.e. at values of
$m^2_3$ corresponding to $T>T_{\rc}$,
one has confinement phenomena like in QCD. There is a linearly
rising part in the potential between static sources \cite{4}
and there exist $W$-ball states almost identical to those in pure SU(2)
YM theory \cite{7},
which fits nicely together with the observation that the string tension 
is approximately the same in both theories. 
Also three-dimensional correlation
masses of $0^+, 1^-, 2^+$ bound states of Higgses have been
``measured'' \cite{7}.
They can be described \cite{8} in a simple relativistic bound
state model analogous to the one of Simonov \cite{9}
in QCD whose main ingredient is an area law for the Wegner-Wilson
loop.

This area law follows naturally from the stochastic vacuum model
of Dosch and Simonov \cite{10,11} which is quite successful in QCD.
It postulates the following
gauge and Lorentz covariant correlation function
between field strengths ${\DF}_{\mu\nu}
\equiv F^a_{\mu\nu}T^a$ at different positions $x$ and $x'$
\bear\label{1.3}
&&\ll g^2F^a_{\mu\nu}(x',x_0)F^b_{\lambda\sigma}
(x,x_0)\gg^{\rm NP}=\frac{\delta^{ab}}{N_c^2-1}\frac{<g^2F^2>}{d(d-1)}\times
\nonumber\\
&&\times \Bigl[(\delta_{\mu\lambda}\delta_{\nu\sigma}-\delta_{\mu\sigma}
\delta_{\nu\lambda})\kappa D(z^2/a^2)\nonumber\\
&&+\frac{(1-\kappa)}{2}(\partial_\nu(z_\sigma\delta_{\mu\lambda}-z_\lambda
\delta_{\mu\sigma})-(\mu\leftrightarrow\nu))D_1(z^2/a^2)\Bigr]\ear
to dominate the cumulant expansion (see eq. (\ref{3.4})).
Here $<g^2F^2>$ is the usual $x$-independent condensate, and $D$
and $D_1$ are form factors, containing a correlation length $a$ 
normalized such that $D(0)=D_1(0)=1$. The way it is used in this 
paper (\ref{1.3}) only contains the nonperturbative correlation in the nontrivial
QCD type vacuum (we skip the index NP in the following).
Thus in the background field formalism with a decomposition
$A_{\mu}^a=A_{\mu}^{a(\rm background)}+a_{\mu}^{a(\rm quantum)}$
it describes a correlated gauge field background. To obtain the
nonperturbative correlation in the continuum limit from lattice
calculations the (diverging) perturbative part has to be subtracted 
(or suppressed by ``cooling''). The quantum field $a_{\mu}^a$ propagates
and couples according to the usual perturbation theory. We expect 
that the propagation of the soft quanta should be suppressed by
the background and this will indeed turn out to be the case.

Furthermore
\be\label{1.4}
F^a_{\mu\nu}(x,x_0)=\left[P\exp(ig\int^x_{x_0}d\bar x_\lambda
A_\lambda(\bar x))\right]^{ab}F^b_{\mu\nu}(x)\ee
is the field strength tensor parallel-transported
to a fixed reference point $x_0$
by a Schwinger string in the adjoint representation.
The integration path is fixed to be a straight line. But even then the
l.h.s. of (\ref{1.3}) generically depends on $x_0$ whereas the r.h.s.
only depends on $z=x'-x$. Neglect of the $x_0$ dependence is
an assumption \cite{11}, which is fulfilled in an appropriate range 
of $x_0$ where the cumulant expansion converges
rapidly: The Wegner-Wilson loop vev is $x_0$ independent
and if it is evaluated with stochastic correlations also these
have to be $x_0$ independent.
 With a linear path in (\ref{1.4}) and with $x_0$ on the line
$x-x'$ the ansatz (\ref{1.3})
has been tested in 4-dimensional QCD lattice calculations \cite{12}
(and is being also probed in recent 3-dimensional lattice
gauge theory
investigations \cite{13}).

In a coordinate (Fock-Schwinger)
gauge with reference point at $x_0$ the parallel
transport operator in (\ref{1.4}), along a straight
line, equals $\delta^{ab}$. Using this gauge in the correlator (\ref{1.3})
(or more general the nonabelian Stokes theorem)
in order to perform a cumulant expansion of the
vacuum expectation value of the Wegner-Wilson loop with the
higher cumulants neglected, one easily obtains \cite{10,11}
the area law. The
string tension is related to the local condensate $<g^2F^2>$ and
to the correlation length $a$
which enters  the form factors $D$ and  $D_1$ (see eq. (\ref{3.6})).

Applying this picture to the three-dimensional
Higgs-gauge theory (\ref{1.1}), a stochastic
gauge field background would not only
be present in the ``hot'' phase, but more generally at
small Higgs background fields $\phi$.
It leads to a modified effective potential containing nonperturbative
effects of the gauge field background. This has
been already proposed \cite{13a} some time  ago
for constant $<F^2>$. In the present
paper the importance of nontrivial form factors describing
a correlated gauge field background is demonstrated.

We propose that the nonperturbative dynamics of the theory (1.1)
is dominated by a fluctuating IR-gauge field background like in
pure YM theory but with the Higgs field background as a 
further parameter. Our most important starting point will be the 
instability of a vacuum with $<F^2>=0$ for small background
Higgs fields. Such an instability we obtain due to paramagnetic
interactions related to the spin of the ``W-gluons''. It is
similar to the Savvidy instability of QCD.   But it is now stated
for fluctuating fields, and there are no IR singularities.

Chapter 2 contains a discussion of the
order $F^2$-term in an effective potential $V$ depending 
on the gauge and Higgs fields and of the instability
of a vacuum with $<F^2>=0$. In chapter 3 we calculate the 
1-loop effective
potential in a certain approximation. We introduce a momentum 
dependent positive $(\mbox{mass})^2$ $m^2_{conf}$ of the
gauge boson  due to confinement and a negative momentum dependent
$(\mbox{mass})^2$ $-\Sigma$ related to paramagnetic spin-spin
interactions. The evaluation of the potential in chapter 4 allows us to
qualitatively discuss the crossover behavior of the hot electroweak
theory. Appendix A contains a thorough discussion of the gluon 
propagator in the stochastic vacuum background. While the
basic idea of how stochastic background fields can lead to
the formation of condensates is very easy to understand, its
implementation in a gauge theory is rather involved.
Therefore, in Appendix B, we discuss this mechanism for the
much more transparent case of a simple scalar theory. The reader
might wish to turn to this Appendix before embarking on the 
detailed presentation in section 2 and 3.

\section{\protect\mbox{Instability of $F^2=0$ for small Higgs
field $\varphi$}}
\setcounter{equation}{0}

As a first step let us calculate the term
\be\label{2.1}
\ll \Gamma_{FF}\gg=\int d^3x V_{FF}(\varphi)\ee
 in the 3-dimensional effective action which is
{\it linear} in the condensate\\ $<g^2_3F^2>
\equiv<g_3^2F_{\mu\nu}^aF^a_{\mu\nu}>$:
\be\label{2.2}
V_{FF}(\varphi)=<g^2_3F^2>P(\varphi^2)\ee
where $\varphi=\sqrt{2<\phi^+\phi>}$.
The potential $V_{FF}$ arises from the stochastic average
of that term in the ordinary effective action $\Gamma[A,\varphi]$
which contains two powers of
$F_{\mu\nu}$ and an arbitrary number of covariant derivatives
acting on them:
\begin{eqnarray}\label{2.3}
\ll\Gamma[A,\varphi]\gg&=&\ll-\frac{1}{4} g^2\int d^dx \mbox{ tr}_c  
{\bf F}_{\mu\nu}(x)\Pi(-D^2){\bf F}_{\mu\nu}(x)+...\gg \\
&=&\ll\Gamma_{FF}\gg+... \nonumber
\end{eqnarray}
Here $\Pi$ is the $\varphi$-dependent polarization function (divided
by $g^2$) and $\mbox{ tr}_c$ denotes a color trace in the adjoint
representation. In order to be slightly more general, we shall
consider a $SU(N_c)$ gauge theory in $d$ dimensions and set $N_c=2,\ d=3,
\ g\equiv g_3$ only at the very end. To proceed, we evaluate the
gauge-invariant action (\ref{2.3}) for gauge fields which satisfy
the Fock-Schwinger gauge condition
\be\label{2.4}
(x-x_0)_\mu A^a_\mu(x)=0\ee
for some arbitrary point $x_0$. Gauge fields satisfying (\ref{2.4})
can be expressed in terms of the corresponding field strength according
to
\be\label{2.5}
A^a_\mu(x)=\int^1_0d\eta\eta(x-x_0)_\nu F^a_{\nu\mu}(x_0+\eta(x-x_0))\ee
As a consequence, there is a one-to-one correspondence
between powers of $A_\mu$ and powers of $F_{\mu\nu}$, and
being interested in $F^2$-terms only we may ignore the $A_\mu$-terms
in the covariant derivatives of (\ref{2.3})
\be\label{2.6}
\Gamma[A,\varphi]=\frac{1}{4}g^2N_c\int d^dx
F_{\mu\nu}^a(x)\Pi(-\partial^2)F^a_{\mu\nu}
(x)+...\ee

A priori the reference point $x_0$ in the stochastic correlator (\ref{1.3})
is unrelated to the point $x_0$ above which characterizes a specific
Fock-Schwinger gauge. In the following we shall identify these two
points. This has the consequence that the parallel-transport operator
in eq. (\ref{1.4}) becomes equal to the unit matrix. Only in this case
we are dealing with the correlator of two local field strength operators.
In particular,
\be\label{2.7}
\ll g^2F^a_{\mu\nu}(x')F^a_{\mu\nu}(x)\gg=<g^2F^2>D_{\rm eff}
\left(\frac{z^2}{a^2}\right)\ee
with the abbreviation
\be\label{2.8}
D_{\rm eff}\left(\frac{z^2}{a^2}\right)\equiv\kappa D
\left(\frac{z^2}{a^2}\right)+(1-\kappa)D_1\left(
\frac{z^2}{a^2}\right)+(1-\kappa)
\frac{2z^2}{da^2}D_1'
\left(\frac{z^2}{a^2}\right)\ee
Now it is straightforward to compute 
$\ll \Gamma_{FF}\gg$ by applying (\ref{2.7}) to (\ref{2.6}):
\be\label{2.9}
V_{FF}(\varphi)=\frac{1}{4}N_c<g^2F^2>\int\frac{d^dp}{(2\pi)^d}
\Pi(p^2)\tilde D_{\rm eff}(p^2)\ee
Here $\tilde D_{\rm eff}(p^2)$ is the Fourier
transform of $D_{\rm eff}(z^2/a^2)$.
Denoting the Fourier transforms of $D(z^2/a^2)$ and $D_1(z^2/a^2)$ as
$\tilde D(p^2)$ and $\tilde D_1(p^2)$, respectively, it reads
\be\label{2.10}
\tilde D_{\rm eff}(p^2)=\kappa\tilde D(p^2)-\frac{2}{d}(1-\kappa)p^2
\tilde D_1'(p^2)\ee

Later on in our numerical computations we shall consider the case
$\kappa=1$ only\footnote{The $D_1$ form factor does not contribute 
to the area law \cite{10,11}} and use  the two form factors
\be\label{2.10a}
D^{(1)}\left(\frac{z^2}{a^2}\right)=e^{-|z|/a}\ee
\be\label{2.10b}
D^{(2)}\left(\frac{z^2}{a^2}\right)=e^{-z^2/a^2}\ee
Their Fourier-transforms for $d=3$ are, respectively,
\bear
\tilde D^{(1)}(p^2)&=&\int d^3ze^{ipz}e^{-|z|/a}
=\frac{8\pi a^3}{(1+a^2p^2)^2}\label{2.10c}\\
\tilde D^{(2)}(p^2)&=&
\pi^{3/2}a^3\exp\left(-\frac{a^2}{4}p^2\right)\label{2.10d}.\ear

It remains to compute the polarization function $\Pi$. We
restrict ourselves to a one-loop calculation here. The dominant
contributions to $\Pi$ come from gauge boson and ghost loops.
(Higgs loops play a minor role and are considered later on.)
Using the proper time method, their contribution to the effective
action is given by
\bear\label{2.11}
\Gamma&=&\frac{1}{2}{\rm Tr}{\rm ln}[K]-{\rm Tr}
{\ln}[-D^2+m^2]\nonumber\\
&=&-\int^\infty_0\frac{dT}{T}\left\{\frac{1}{2}
{\rm Tr} [e^{-TK}]-{\rm Tr}[e^{-T(-D^2+m^2)}]\right\}\ear
Here
\be\label{2.12}
K^{ab}_{\mu\nu}=\left[-D^2\delta_{\mu\nu}+2ig F_{\mu\nu}+m^2
\delta_{\mu\nu}\right]^{ab}
\ee
is the kinetic operator of the gauge bosons in the Feynman-'t
Hooft gauge and $-D^2+m^2$ the
corresponding one for the ghosts. 
We identify
\be\label{2.13}
m^2\equiv\frac{1}{4}g^2_3\varphi^2,\quad\varphi=\sqrt{2<\phi^+\phi>}\ee
Using the method of Barvinsky and Vilkovisky \cite{13b}
or the corresponding world line technique \cite{13c}
it is straightforward to identify the $F^2$-terms which are contained
in the functional traces of (\ref{2.11}):
\bear\label{2.14}
\Pi(p^2)&=&-\frac{4}{(4\pi)^{d/2}}\int^\infty_0dTe^{-m^2T}T^{-d/2+1}
\nonumber\\
&&\cdot\left[f(p^2T)+\frac{1}{4}(d-2)\frac{f(p^2T)-1}{p^2T}\right]
\ear
with the familiar ``second-order form factor''
\be\label{2.15}
f(p^2T)\equiv \int^1_0d\alpha\exp[-\alpha(1-\alpha)p^2T]\ee
Performing the proper time integral yields $\Pi=\Pi_G+\Pi_F$
with
\bear\label{2.16}
&&\Pi_G(p^2)=-\frac{(d-2)\Gamma(1-d/2)}{(4\pi)^{d/2}}\frac{1}{p^2}\int
^1_0d\alpha\left[(m^2+\alpha(1-\alpha)p^2)^{d/2-1}-m^{d-2}
\right]\nonumber\\
&&\Pi_F(p^2)=-\frac{4\Gamma(2-d/2)}{(4\pi)^{d/2}}
\int^1_0d\alpha\left[m^2+\alpha(1-\alpha)p^2\right]^{d/2-2}\ear
The contribution $\Pi_F$ originates from the nonminimal
coupling of the gauge boson fluctuations to the background
(the $F_{\mu\nu}$-term in $K_{\mu\nu}$), while $\Pi_G$
arises from the $D^2$-term in $K_{\mu\nu}$ and from the ghosts.
Using (\ref{2.16}) in (\ref{2.9}) for $d=3$ and $N_c=2$ we obtain
the desired expression for the potential:
\bear\label{2.17}
&&V_{FF}(\varphi)=-\frac{1}{8\pi^3}<g^2_3F^2>\int^\infty_0
dp\ p^2\tilde D_{\rm eff}(p^2)\\
&&\cdot\int^1_0d\alpha\left\{\left[m^2+\alpha(1-\alpha)p^2\right]
^{-1/2}-\frac{1}{2p^2}\left[(m^2+\alpha(1-\alpha)p^2)^{1/2}-m
\right]\right\}
\nonumber\ear
The momentum integration in (\ref{2.17}) converges both for $p\to\infty$
and for $p\to0$. In fact, $V_{FF}$ is IR-finite even in the limit
$m\equiv g_3\varphi/2\to0$ of massless gauge bosons and ghosts.
We emphasize that this IR finiteness has nothing to do with the
dynamical generation of a mass for these particles (see below),
but rather with the ``inclusive'' nature of the quantity at hand.
For $d=3$ and $m=0$ the polarization function
\be\label{2.18}
\Pi(p^2)=-\frac{1}{2}(1-\frac{1}{16})\frac{1}{\sqrt{p^2}}\ee
is singular at $p=0$, of course, but the volume element
$d^3p=4\pi p^2 dp$ in (\ref{2.9}) has the effect of completely
suppressing the contribution of $p=0$ to $V_{FF}(\varphi)$.
($\tilde D_{\rm eff}(0)$ is finite and nonvanishing.)

In eq. (\ref{2.18}) the ``$1$'' comes from $\Pi_F$ while the
``$-1/16$'' is due to $\Pi_G$. We observe that the nonminimal
$F_{\mu\nu}$ interaction of the gauge boson fluctuations
is the dominant effect. The contribution from the
minimal couplings of gauge bosons and ghosts is smaller
by more than one order of magnitude and has the opposite
sign. This situation persists also for $m>0$.

It is important to note that the one-loop
contribution $V_{FF}(\varphi)$ is negative.
It has to be added to the stochastic average
of the positive tree-level
term,
\[\ll\frac{1}{4}F^a_{\mu\nu}F^a_{\mu\nu}\gg=<g_3^2F^2>/4g^2_3.\]
If it is large enough so that the sum is still negative, this
indicates that the vacuum with a vanishing gauge field condensate is
unstable and that the true ground state of the system is
characterized by a nonvanishing value of $<F^2>$.\footnote{See
ref. \cite{gluco} for a related discussion in the exact renormalization
group framework.}

At the linearized level, the dynamics of the gauge boson fluctuations
$a_\mu(x)$ is governed by the Lagrangian
\be\label{3.6a}
{\cal L}=\frac{1}{2}a_\mu\left[-D^2\delta_{\mu\nu}+2ig
F_{\mu\nu}+m^2\delta_{\mu\nu}\right]a_\nu+O(a^3_\mu)\ee
The coupling $a_\mu F_{\mu\nu}a_\nu$ is the spin-1 analogue
of $\bar\psi\vec\sigma\cdot\vec B\psi$ for spin-$\frac{1}{2}$
particles. It gives rise to a ``paramagnetic'' interaction of the
gauge bosons, while the terms resulting from the minimal substitution,
$a_\mu D^2a_\mu$ lead to the more conventional ``diamagnetic'' behavior  
\cite{huang}. Thus, using the terminology of solid state physics,
we can say that the instability found above results from a
dominance of paramagnetic over diamagnetic effects.

It is interesting to note that the possible existence of such an
instability is closely related to the fact that the gauge bosons
carry spin 1 and have a (tree level) Land\'e factor $g_L=2$. This
is most easily seen if we expose a ``colored'' particle of
spin $S=\frac{1}{2}$ or $S=1$ with $g_L=2$ to a covariantly constant
color magnetic field of strength $B$ along the $z$-axis. Its
squared one-particle energies are given by \cite{huang, dr}
\be\label{3.6b}
E^2=m^2+p^2_z+gB(2n+1)-2gBS_z\ee
where $n=0,1,2,...$ enumerates the Landau levels and $S_z$ is the
spin projection along the $z$-axis. In eq. (\ref{3.6b}) the
terms $gB(2n+1)$ and $-2gBS_z$ stem from the diamagnetic and
the paramagnetic interactions, respectively. An instability
is signalled by an imaginary part of $E$. In the case of fermions
we have the two possibilities $S_z=-1/2$ and $S_z=+1/2$. In
the former case, $E^2$ is strictly positive for all $n$, but in the
latter the paramagnetic term precisely cancels the diamagnetic
one at $n=0$. For $m^2=0$ one finds at $p_z=0$ a marginally
stable state at $E^2=0$. For a spin-1 particle one has $S_z=0,\pm1$.
Hence, for the spin properly aligned to the external
magnetic field, the  diamagnetic energy of the lowest Landau
level, $+gB$, combines with a paramagnetic contribution $-2gB$ to
yield a negative total magnetic energy $-gB$. Hence $E^2<0$ for
$m^2+p_z^2$ sufficiently small. (This effect is responsible
for the instability discussed in the context of the Savvidy 
vacuum of QCD, for instance.)

The effective action which we have calculated above describes the
interaction of gauge boson fluctuations with non-constant external
fields of arbitrary momentum. Hence their spectrum is certainly
not given by (\ref{3.6b}). However, what continues to be true is
that for the spin-1 field the paramagnetic interaction overrides
the diamagnetic one -- which would not be possible for Dirac
fermions.

When $\varphi^2$ is increased from $\varphi^2=0$ to larger values,
the potential $V_{FF}(\varphi)$ approaches zero from below.
Thus, starting from $<F^2>\not=0$ at $\varphi^2=0$, one will
return to a situation with vanishing condensate for $\varphi^2$
larger than a certain critical value $\varphi^2_c$.

In Appendix B we discuss a scalar toy model
and give a simple physical explanation of why stochastic
background fluctuations tend to trigger the formation of
condensates.

Taking $a\to\infty$ in eq. (\ref{2.17}) with form factors (\ref{2.10c})
or (\ref{2.10d}) leads back to the perturbative result
$\sim g^2_3<F^2>/m$,
implying the 1-loop gauge $Z$-factor\footnote{There is a misprint
in formula (A.12) of ref. \cite{kls}.}
\be\label{2.19}
Z_{\rm gauge}=1-\frac{g^2_3}{8\pi}
\left[\frac{7}{m_W}+\frac{2}{3m_{gh}}-\frac{1}
{8m_{Gb}}-\frac{1}{24 m_H}\right]\ee
with
\[m_W^2=m^2_{gh}=\frac{1}{4}g^2_3\varphi^2;\quad
m^2_{Gb}=\frac{1}{4}g^2_3\varphi^2+\lambda_3\varphi^2
+m^2_3;\quad m^2_H=3\lambda_3\varphi^2+m^2_3\]
It has an IR singularity at $m^2=\frac{1}{4}
g^2_3\varphi^2=0$ which is removed in our calculation with
a gauge field correlator. Indeed for $a\to\infty$ we obtain
the $W$-boson plus ghost part of $V_{FF}$ as $-({23}/{12\pi})
g_3^2\left(\frac{1}{4}F^2\right)$
as given in eq. (\ref{2.19}).

The scalar Higgs contributions in the loop can be also taken
into account in a Barvinsky-Vilkovisky type formula. Higgs
fields in the loop are easily included in $V_{FF}$
by additional terms  in $\Pi$ proportional to
$(f(-T\partial^2)-1)/(-T\partial^2)$, and by modified
prefactors $\frac{1}{8N_c}e^{-m^2_HT}$ and $\frac{3}
{8N_c}e^{-m^2_{Gb}T}$ for the
Higgs boson and the Goldstone bosons,
respectively,  instead of $(d-2)e^{-m^2T}$.
This is in agreement with (\ref{2.19}) for $a\to \infty\ (N_c=2, d=3)$.
It is obvious that these parts are small compared to the gauge
parts.
Mixed gauge boson-Higgs field graphs do not contribute. 

\begin{figure}[t] 
\begin{picture}(200,210)
\put(95,0){\epsfxsize9cm \epsffile{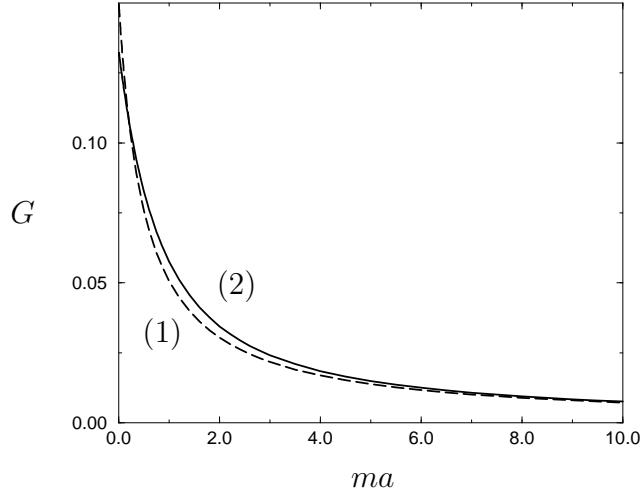}}
\put(220,10){ $ma$}
\put(95,110){$G$}
\put(145,65){(1)}
\put(173,83){(2)}
\end{picture} 
\caption{The function $G(ma)$ appearing in $V_{FF}$ of eq. (2.26)
for the two different form factors}
\label{Fs} 
\end{figure}

Evaluation of $V_{FF}$ in eq. (\ref{2.9}) for $\kappa=1$, i.e. with
a pure $D$-form factor gives
\be\label{2.23}
V_{FF}=-<g^2_3F^2>aG(ma)\ee
with
\be\label{2.24}
G(ma)=\frac{(ma)^2}{16\pi^3}\int^\infty_0d\bar p\bar D(m^2a^2\bar p
^2)\left[\left(\frac{15}{4}\bar p-\frac{1}{\bar p}\right)
\left(\arcsin \frac{\bar p}{\sqrt{\bar p^2+4}}
-\frac{\bar p}{2}\right)+\frac{15}{8}\bar p^2\right]\ee
where $\tilde D(p^2)\equiv a^3\bar D(m^2a^2\bar p^2)$ with
$\bar p\equiv p/m$.
For $m^2\to0$ this simplifies to
\be\label{2.25}
G(0)=\frac{15}{128\pi^2}\int^\infty_0dp
p\bar D(p^2)\ee
The first term inside the curly brackets of
eq. (\ref{2.17}) derives from
the $F_{\mu\nu}$ part in $K_{\mu\nu}$ of eq. (\ref{2.12}).
It contributes $4\bar p \arcsin (...)$ in the
bracket in (\ref{2.24}) and dominates $V_{FF}$. It is
responsible for the negative sign of $V_{FF}$ which
is capable of generating an instability at $F^2=0$.

The numerical evaluation of (\ref{2.24}) for\\

 1)
$\bar D^{(1)}(k^2)=8\pi/(1+k^2)^2$ and \\

 2)
 $\bar D^{(2)}(k^2)=\pi^{3/2}\exp(-k^2/4)$ \\
\mbox{ } \newline
corresponding to the form factors (\ref{2.10a}) and (\ref{2.10b}),
respectively,
is shown in fig. \ref{Fs}.
The massless limit is given by
(\ref{2.25}). $G(0)$ of eq. (\ref{2.24})
is calculated as
\begin{figure}[t]
\begin{picture}(200,110)
\put(100,20){\epsfxsize8cm \epsffile{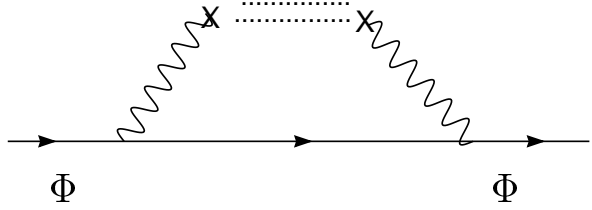}}
\end{picture}
\caption{Higgs field interaction with stochastic gauge
field background}
\label{treeterm}
\end{figure}
\be\label{2.26}
G^{(1)}(0)\simeq0.1492\ee
and
\be\label{2.27}
G^{(2)}(0)\simeq0.1322\ee
respectively. Note that defining $\bar a=\int^\infty_0dzD(z^2)$
one has $\bar a=a$ in case (1) and  $\bar a=0.85a$ in case (2); thus
$aG(0)$ has a very similar value in both cases if expressed
in terms of $\bar a$.

There is a further contribution to $V_{FF}$ by a Higgs field
tree term (fig. \ref{treeterm}). Because of the correlator it is similar
to a 1-loop term. The gauge-boson coupling to an
external constant Higgs
field is given by $(-ig\partial_\mu A^a_\mu)(\Phi^+T^a\varphi_{\rm qu}-
\varphi^+_{\rm qu}T^a\Phi)$, where $\varphi_{\rm qu}$ is the
quantum fluctuation.
The correlator
\be\label{2.20}
\ll\partial_\mu A^a_\mu(x)\partial_{\bar\mu}A^b_{\bar\mu}(x')\gg\ee
can be conveniently evaluated in the Fock-Schwinger (coordinate)
gauge (\ref{2.5}) (like the Barvinsky-Vilkovisky form factors \cite{13b})
with $x_0=\frac{x+x'}{2}$ (since we are discussing the contribution
to an effective action with quasilocal terms). Note that the
differentiation in (\ref{2.20}) should not affect the $x,x'$ in $x_0$.
Fourier transformation to  momentum space gives
\bear\label{2.21}
&&V_{FF}^{\rm tree}=\frac{2}{9}
<g^2_3F^2>\Phi^+T^aT^a\Phi\frac{1}{(2\pi)^2}\int^\infty_0
dqq\int^1_0d\eta d\bar\eta\frac{\eta\bar\eta}{(\eta+\bar\eta)^5}\\
&&\lim_{p\to 0}\left(\frac{2}{p}\log
\frac{(p+q)^2+m^2_3}{(p-q)^2+m_3^2}\right)\left(3\tilde D
\left(\frac{4q^2}{(\eta+\bar\eta)^2}\right)+\frac{8q^2}{(\eta
+\bar\eta)^2}{\tilde D}'\left(\frac{4q^2}{(\eta+\bar\eta)^2}
\right)\right)\nonumber\ear
The limit $\lim_{p\to 0}(...)$ equals  $8q/(q^2+m^2_3)$. After an
integration by parts of the second term with
respect to $q$, the $\eta-\bar\eta$
integration can be done analytically. The $(\eta+\bar\eta)/2$
and $q$ integrations were done numerically for the special form
factors (\ref{2.10c}) and (\ref{2.10d}). 
\begin{figure}[t] 
\begin{picture}(200,210)
\put(95,-10){\epsfxsize9cm \epsffile{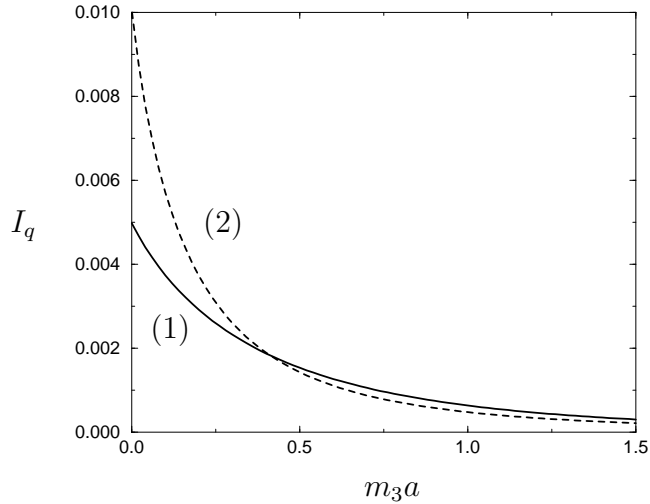}}
\put(220,0){ $m_3a$}
\put(90,100){$I_q$}
\put(143,60){(1)}
\put(163,100){(2)}
\end{picture} 
\caption{$I_q(m_3a)$ contributing to $V_{FF}^{tree}$ in (2.33) for the 
two form factors (2.11) and (2.12) indicated by (1) resp. (2)}
\label{Iq}
\end{figure}
This results in
\be\label{2.22}
V^{\rm tree}_{FF}=\frac{\varphi^2}{4}a^2I_q(m_3a)
<g^2_3F^2>\frac{1}{6\cdot3}\ee
where
\[I_q(m_3a)=\frac{6}{(2\pi)^3}\int^\infty_0d\bar q\frac{
\bar q^4}{(\bar q^2+a^2m^2_3)^2}I_\eta(\bar q)\]
with some function $I_\eta(qa)$ which can be extracted from (\ref{2.21}).
The function $I_q(m_3a)$ is plotted in fig. \ref{Iq}. Note that $I_q(m_3a)$ 
does not exist for $m^2_3<0$. $I_q$ is a monotonically
decreasing function of $am_3$ with
$I_q(0)\approx0.005-0.01$. $V_{FF}^{\rm tree}$ is {\it positive}
and thus supports the vanishing of the gauge condensate for increasing
$\varphi^2$.

The potential $V_{FF}$ of eq. (\ref{2.23})
and the additional small positive pieces including the $V_{FF}^{\rm tree}$
discussed above
have to be added to the tree potential
$<\frac{1}{4}F^2>$. From eq. (\ref{2.23}) and (\ref{2.22}) we obtain an
instability at $F^2=0$ if
\be\label{2.28}
g^2_3aG(ma)-m^2a^2I_q(m_3a)/18>0.25\ee
An instability at $F^2=0$ requires a
correlation mass $1/a<4G(0)g^2_3
\approx 0.6g^2_3$.

Lattice studies \cite{13}
indicate $1/a\sim0.73 g^2_3\sim m_{\rm glueball}/2$,
as suggested in ref. \cite{8}. Strictly speaking, our calculation
does not give an instability with such a value of $1/a$.
However, keeping in mind the limitations of the present
semi-quantitative approach, we have
to be aware that it might be not very accurate.
It is also gauge dependent. In perturbative
calculations of the electroweak potential one observes that a 
strong gauge dependence of the 1-loop order result indicates 
that the next loop order contribution is needed and 
that one can arrive in 2-loop order at a quantitative reliable result
without much gauge dependence \cite{1,6}.
Thus the 
gauge dependence of the 1-loop expression is a very useful hint.
A remaining gauge dependence of the potential \cite{6}
together with a similar one of the $Z$-factor in the
kinetic term should drop out in the calculation of the
action of an extremal field configuration such as the 
critical bubble. But this has not really been possible
to check up to now 
even in pure perturbative calculations. 
Here we employed the Feynman-'t Hooft gauge. 
In our present treatment the 
use of a specific model for the IR sector of the theory
could weaken, but certainly not erase the gauge-fixing dependence.
A 2-loop calculation in the perturbative part is very
demanding, but would be very interesting.

A negative $F^2$ term then leads to a ground state
with $<F^2>\neq 0$ to be stabilized by confinement
forces in an effective potential $V(\varphi^2,<g_3^2F^2>)$
as we will argue in the following chapters. 
The gauge boson condensate 
is determined
as the minimum position of the ``potential'' in $F^2$. Given the minimum
value of this potential, its negative linear part at small $F^2$ allows
a rough estimate of the minimum position (e.~g.~allowing for an
$F^4$ term). Thus the lattice determination of the nonperturbative
finite part $a_4g_3^2$ of the free energy in ref. \cite{klpr} in pure 
Yang-Mills theory $(m^2=0)$ may be
used to connect the $F^2$ coefficient discussed in section 2
with the gauge condensate and  with the string tension also known from
the lattice. The small negative $a_4$ reported in \cite{klpr} then points to
a negative coefficient of $\frac{1}{4}F^2$ with an absolute value much 
smaller than $1$. This strengthens our view that we are close to
instability with the lattice value of $a$ and that only small further
(2-loop) contributions are needed to get complete agreement. 
A careful analysis including lattice renormalization is required.

The instability of $<F^2>=0$ is strengthened with growing
$g_3^2$. If we define the IR running of $g^2_{3_{\rm eff}}$ via
the prefactor of the $F^2$ term in the effective action,
the negative sign of the 1-loop contribution tells us that
$g^2_{3_{\rm eff}}$ is indeed larger than $g^2_3$ and that it
would diverge at the border of instability of $<F^2>=0$.
In the Wilsonian renormalization-group approach of ref.
\cite{2a} $g^2_{3_{\rm eff}}(k)$ at a running IR scale $k$
is the central quantity. Its divergence at some $k=k_c$
signals confinement like in QCD. It is obtained from
a differential renormalization-group procedure and certainly
goes beyond a 1-loop order calculation. One way to look at the
effective action (average action) of this approach is to
supplement the usual
background field formalism with a smooth IR cut-off $k$ for the
momenta of inner propagators \cite{W}. The stochastic vacuum
with a correlation length $a$ (together with the
condensate $<F^2>$, both in principle to be determined
dynamically) then constitute some ``physical'' value for
the IR scale $k$.

\section{The effective action with stochastic\protect\\ vacuum
 correlations}
\setcounter{equation}{0}

In the previous section, only those terms in $<<\Gamma[A,\varphi]>>$
were calculated which are linear in $<g^2_3F^2>$. Let us now turn
to the much more complicated question of how $<<\Gamma>>$
depends on $<g^2_3F^2>$ to all orders.

The 1-loop effective action with a gauge boson circulating
in the loop and
with a Higgs field plus a correlated gauge field
background can be easily written
down as a worldline path integral:
\bear
&&\Gamma_{\rm gauge}(m^2,<g_3^2F^2>)=-\frac{1}{2}\int^\infty_0 \frac{dT}{T}
\int_{x(0)=x(T)}[Dx]\exp\left\{-\int^T_0d\tau\left(
\frac{\dot x_\mu^2(\tau)}{4}+m^2\right)\right\}\nonumber \\
&&\ll {\rm tr}_{\rm cL}P\exp\left\{-\int^T_0d\tau(-ig)
\left({\bf A}_\mu(x)
\dot x_\mu\eins_{\rm L}+2{\DF}(x)\right)\right\}\gg\label{3.1}\ear
The subscripts ``c'' and ``L'' refer to the color and Lorentz
matrix structure, respectively.
We use the coordinate gauge based at the center of mass
point $x_0$ \cite{13c}
defined by
\be\label{3.2}
x(\tau)=x_0+y(\tau),\quad \int^T_0d\tau y(\tau)=0,
\quad Dx=d^dx_0Dy\ee
which amounts to
\be\label{3.3}
A^a_\mu(x_0+y)=\int^1_0d\eta\ \eta\ y_\nu F^a_{\nu\mu}(x_0+\eta y)\ee
The parallel transport implicit in the correlator on the l.h.s.
of (\ref{1.3}) is taken along a $V$-shaped path consisting
of two straight lines from $x$ to $x_0$ and from $x_0$ to $x'$,
respectively.
The path is of course not
the straight line connecting $x$ to $x'$ which is normally used 
in lattice calculations
\cite{12}. One has to
make the assumption that the correlation length is independent
of the position $x_0$ somewhere inside the loop \cite{11}.\footnote{Some
indication that the correlation length is path independent can be
found in ref. \cite{ev}} 

The average over the
gauge field background $\ll..\gg$ in
(\ref{3.1}) can be evaluated using the lowest (quadratic) term
in the cumulant expansion. This is the ``stochastic vacuum''
model assumption
\be\label{3.4}
\ll e^{\int A}\gg\approx \exp(-1/2\int\int\ll AA\gg)\ee
The correlator in the exponential follows from eq. (\ref{1.3})
with (\ref{3.3}).
After rescaling $\tau=T\bar\tau,\ y=T^{1/2}\bar y$ and taking
only the form factor $D$ in (\ref{1.3}) $(\kappa=1)$, we obtain
for the correlated part in (\ref{3.1})
\bear\label{3.5}
&&\Gamma_{\rm gauge}(m^2,<g_3^2F^2>)=-\frac{1}{2}\int^\infty_0\frac{dT}{T}
T^{-d/2}\int d^dx_0\int[D\bar y]\exp\{-\int^1_0
d\bar\tau\frac{\dot{\bar y}^2}{4}-m^2T\}
\nonumber\\
&&\times {\rm tr}\ \exp \left[-\frac{<g^2_3F^aF^a>}{d(d-1)}
\frac{{\bf \eins}_{\rm c}}{(N_c^2-1)}T^2\int^1_0d\bar\tau d\bar
\tau'\right.\nonumber\\
&&\left\{\int^1_0d\eta d\eta'(\eta\eta')(-(
\bar y\cdot\dot{\bar y}')^2+
\bar y\cdot\bar y'\dot{\bar y}\cdot\dot{\bar y}')
\eins_{\rm L}D\left(\frac{(\eta\bar y-\eta'\bar y')^2T}
{a^2}\right)\right.
\nonumber\\
&&+4(1-d)\eins_{\rm L}D\left(\frac{(\bar y-\bar y')^2T}{a^2}\right)\\
&&\left.\left.+\int^1_0d\eta\eta(\bar y\times \dot{\bar y}'-\dot{\bar y}
\times {\bar y}')D\left(\frac{(\eta\bar y-\bar y')^2T}{a^2}\right)
+({\rm primed}\leftrightarrow {\rm unprimed})\right\}\right]\nonumber\ear
Expansion to order $F^2$ would reproduce the contribution from gauge bosons
in the loop calculated in chapter 2.

The ghost contribution  $\Gamma_{\rm ghost}$ has the opposite
sign and an additional factor of 2 compared
to (\ref{3.1}). The trace is only over adjoint color indices, and only
the first term in the curly brackets in (\ref{3.5}) appears.

The first term of (\ref{3.5})
originates from the correlation of two Wegner-Wilson loop
integrals $\exp(-\oint Adx)$. Using the nonabelian Stokes
theorem one obtains area integrals and the correlator
(\ref{1.3}) between two $F$'s for large $(d\gg a)$ loops leads to one common
area integral. This is a nice and well-known result of the stochastic
vacuum model \cite{10,11}. The first term in (\ref{3.5})
is just this area
integral. For large areas it reduces to
$\bar\sigma\bar A$ with the (rescaled) string tension
\be\label{3.6}
\bar\sigma=\pi<g^2_3F^aF^a>\frac{1}{d(d-1)}
\frac{N_c}{N_c^2-1}T\int^\infty_0
dzzD\left(\frac{z^2}{a^2}\right)=\sigma_{\rm adj}T=
\frac{2N_c^2}{N_c^2-1}\sigma_{\rm fund}T\ee
and $\bar A\equiv A/T$ where $A$ is the area of the (minimal) surface
whose boundary is given by the closed path $y(\tau)$.
In eq. (\ref{3.6}) we wrote
$\sigma_{\rm adj}$ and  $\sigma_{\rm fund}$ for  the
confining string tension in
the adjoint and fundamental
representation, respectively.

The second term in the curly brackets in (\ref{3.5}) arises from the
correlation of two explicit factors of
${\bf F}_{\mu\nu}$ in the exponential of eq. (\ref{3.1}).
It is related
to a gauge boson spin interaction which is short-ranged.
Its range is of the order of the correlation length $a$.
This interaction owes its existence to the nonminimal
``paramagnetic''
coupling of the gauge boson fluctuations $a_\mu(x)$
to the background
field, i.e. to the $F_{\mu\nu}$-term in the kinetic operator
$K_{\mu\nu}$.
The third term in (\ref{3.5}) is due the interference of the
confinement and spin effects. 

To start, let us first disregard the ``diamagnetic'' interactions
coming from the $D^2$-term in $K_{\mu\nu}$ and let us
focus on the spin interaction.
For large loops $(d\gg a)$ the
$\tau,\tau'$-correlations are well approximated by nearest
neighbour interactions along the closed path. Instead of doing the
above world-line
path integral one can change to the
ordinary QFT language and calculate
the gluon self-energy $\Sigma$ in a correlated gauge background.
This calculation is described in Appendix A in
some detail. There we also include the
covariant derivative induced parts of the interaction. Of course,
the use of the correlator (\ref{1.3}) again requires the
specification of a reference point for the parallel transport operator
and for the coordinate gauge. In the computation of $\Sigma(x,x')$
we identify both of those points with $x_0\equiv
(x+x')/2$.

For the spin-induced part of $\Sigma$ we obtain
\be\label{3.7}
\Sigma(x,x')^{ab}_{\mu\nu}={\cS}_F(z^2=(x-x')^2)
\delta_{\mu\nu}\delta
^{ab}\ee
with
\be\label{3.8}
{\cS}_F(z^2)=<g^2F^2>\frac{4N_c}{d(N_c^2-1)}
H(\frac{z^2}{2})D(\frac{z^2}{a^2})\ee
where $H(\frac{z^2}{2})$ is a massive
$(m^2=g^2_3\varphi^2/4)$ scalar
propagator function.
Fourier transformation leads to
\bear\label{3.9}
\tilde{\cS}_F(p^2)&=&+<g^2F^2>\frac{1}{9\pi^2}
\int^\infty_0dq\ q^2\nonumber\\
&&\times \frac{1}{pq}\ln\frac{(p+q)^2+m^2}
{(p-q)^2+m^2}\tilde D(q^2)\ear
where we have inserted the physical values $d=3$ and
$N_c=2$ now. The contribution ${\cS}_F(z^2)$ is given
by the last term of the expression (\ref{A.30}) for
the complete function ${\cS}(z^2)$. 

Eq. (\ref{3.8})
shows that the range of the spin interaction is
determined by the fall-off behavior of the form
factor $D$. Furthermore it is important to
observe that ${\cS}_F(z^2)$ is a positive function.
The self-energy $\Sigma$ is defined in such a
way that the inverse dressed propagator is
$G^{(0)-1}-\Sigma$ where $G^{(0)}$ is the free
one (see eq. (\ref{A.6})). Hence ${\cS}_F>0$
implies a negative (``tachyonic'') contribution
to the effective squared mass of the gauge boson.
This means that the interaction with the stochastic
background destabilizes the ``empty'' vacuum where
the gauge boson fluctuations have a vanishing
expectation value: They have the tendency to
condense which then gives rise to a nonzero
vacuum expectation value of $F^2_{\mu\nu}$.

This conclusion does not change if we retain the
interactions coming from the $D^2$-term in
$K_{\mu\nu}$, i.e. the $\dot x_\mu A_\mu(x)$-term
in the world line path integral (\ref{3.1})
in a coordinate gauge with $x_0=(x_1+x_2)/2$. The
complete result is given in eq.  (\ref{A.34}).
This delivers part of the gauge interaction in $D^2$ into
$\Sigma$ whereas the main part constitutes
the confining force summarized in the area law term discussed
below.
In eq. (\ref{A.37}) we give the complete result for
the mass shift $\Delta m^2$ of the gauge field fluctuations.
The spin interaction contributes the ``1''
in the curly brackets of eq. (\ref{A.37}),
whereas the $D^2$-interactions produce the terms
proportional to $\theta_1$ and $\theta_2$. In view of
(\ref{A.38}) the latter are seen to be negligible
at our present level of accuracy.

Inserting the self-energy (\ref{3.9}) in a gauge boson loop leads to
an effective potential
\be\label{3.10}
V_{{\cS}_F}\sim 2\frac{9}{2}\int\frac{d^3p}{(2\pi)^3}\ln[
(p^2+m^2-\tilde{\cS}_F(p^2))]\ee
Here we have multiplied by 2 because of the two different ways to get
a chain of neighbours. This has to be corrected for the 
zeroth and first order terms
in $\tilde{\cS}_F$. $\tilde{\cS}_F(p^2)$ is positive and
contributes negatively in the propagator.
Of course (\ref{3.10}) has to be UV-renormalized but the counter
terms are identical
to those of the case $\tilde{\cS}_F=0$. Thus (\ref{3.10})  reads
more correctly
\bear\label{3.10'}
&&V_{{\cS}_F}=\frac{9}{2}\int\frac{d^3p}{(2\pi)^3}{\Big\{}2\ln
(p^2+m^2-\tilde{\cS}_F(p^2,m^2))-\ln(p^2+m^2)\nonumber\\
&&+\frac{\tilde{\cS}_F(p^2,m^2)}{p^2+m^2}-\frac{m^2}{p^2}-
\ln p^2{\Big\}}\ear
The ghost loop-induced
part has to be added - it does not contain $\tilde{\cS}_F$.

Let us now focus on the other extreme and consider the
approximation of a
pure area law in the IR for the first term in (\ref{3.5}).
Then it behaves as
$\sim\exp(-\sigma\bar AT)$ for
large $\bar AT$. Furthermore, we assume that 
when $\bar{A}T$ becomes smaller this term
is damped as $\sim \exp(-c\bar A^n<g^2_3F^2>T^n(a^2)^{n-2})$
with an exponent $n>2$.
This damping is supposed to set in for 
$\bar{A}T<\tilde{c}a^2$ where 
$\tilde{c}={\cal O}(1)$ is treated as a free parameter.\footnote{Such 
a behavior with $n=2$ was argued for in the
context of the stochastic vacuum model in ref. \cite{24}.
It would lead to a $m^2_{\rm conf}(p^2)$ in (\ref{3.14})
below with a rather weak fall-off in $p^2$
(like $\tilde S_F(p^2)$). The additional piece in
$\tilde S$ in appendix A due to $\eta$-integrals has also
such a behaviour. Thus a contribution with $n=2$
whould be better associated with the $\Sigma(x,x')$
evaluated before.} In the following we will use
the interpolating ansatz with $n=3$
\be\label{3.11}
\sim\exp\left(-\sigma\bar A\frac{T^3}{(\tilde c a^2/\bar A)^2+T^2}
\right)\ee

If we include also an area law for the ghost loop and
renormalize at
$m^2=F^2=0$ we obtain
\begin{eqnarray}\label{3.12}
V^{\rm area}&=&-\frac{(9-6)}{2}\int^\infty_0\frac{dT}{T}
T^{-3/2}\int D\bar y\exp\left\{-\int^1_0d\bar\tau
\dot{\bar y}^2/4\right\}\times\nonumber\\
&&{}\left[\exp\left(\frac{-\sigma\bar AT^3}{(\tilde c \frac{a^2}{\bar A})^2+T^2}
-m^2T\right)-1+m^2T\right]\end{eqnarray}
where $\bar{A}$ is a complicated functional of the interagtion
variable $\bar{y}(\bar{\tau})$: it is the area of the minimal
surface spanned by the loop $\bar{y}$.

Substituting $(4\pi T)^{-3/2}=\int \frac{d^3p}{(2\pi)^3}e^{-p^2T}$
in (\ref{3.12}) as a  procedure to trade
$T$ for a momentum  integration, the $T$ integration
can be performed (numerically if $\tilde c\not=0$) and we arrive at
\bear\label{3.13}
V^{\rm area}&=&-\frac{(9-6)}{2}\int\frac{d^3p}{(2\pi)^3}
(4\pi)^{3/2}\int D\bar y
\exp\{-\int^1_0d\bar\tau\frac{\dot{\bar y}^2}{4}\}\nonumber\\
&&\left[\ln[p^2+m^2+m^2_{\rm conf}(p^2,\bar A,m^2)]-\ln p^2-\frac{m^2}{p^2}
\right]\ear
where $m^2_{\rm conf}$ is a complicated function 
defined by the requirement of
producing the $\ln[p^2+({\rm mass})^2]$ form of the integrand. 
It is easy to convince oneself that this function always exists
and is positive.
It would be $p^2$-independent and
equal to $\sigma\bar A$ if $\tilde c=0$. Thus we have an area-dependent,
momentum-dependent positive $m^2_{\rm conf}$ related
to confinement. It acts as an IR regulator in the same way
as the Higgs mass but it is momentum-dependent.

The functional integral over $\bar y(\bar\tau)$ cannot be performed
in closed form. It could be performed numerically.
Here we make the rather crude approximation that it
is dominated by paths with areas peaked near a mean value $<\bar A>$.
The numerical value of $<\bar A>$ will be treated as a free
parameter. For $\tilde c=2$ and an average value $<\bar A>=2.5$
the function $m^2_{\rm conf}(p^2,<\bar A>,m^2=0)$ is plotted in fig. \ref{SFM}.

The main contribution to $m^2_{\rm conf}(p^2,<\bar A>,m^2)$ comes from 
$T$$<$$\bar{A}$$>$-values smaller than $\sim 10-20 g_3^{-4}$ which
is not very large. With increasing values of $p^2$ even smaller 
$T$$<$$\bar{A}$$>$-values dominate. At very large areas we would expect gluon loop
holes in the area corresponding to the splitting of the confining string
and virtual production of glueballs. In the case of static valence fields
this would lead to a screening and a flattening of the linear rise of the
2-dimensional potential. As recently measured on the lattice this
only happens in the case of very large distances $d\sim 5-10 g_3^{-2}$
much bigger than the inverse glueball mass 
$\sim a/2\sim(2\cdot 0.73g_3^2)^{-1}$ \cite{13}.
The relation between the 2-dimensional static potential and our effective
action is not clear but we would expect that screening is important
only at very large areas $T\bar{A}$. The screening effect corresponds to
higher correlations and is not described by the stochastic vacuum
model with Gaussian correlations.

The spin-induced
$\Sigma$ and the confinement effects have opposite
impact  on the  ``magnetic mass''. In (\ref{3.5}) there is also
the last term which arises from the interference
of the two effects. It changes sign
under the substitution $\bar y(\bar\tau)\Rightarrow\bar y(1-\bar\tau)$
and thus the term would cancel if it is small.
Leaving it aside one can combine the first (confinement) and second
(spin interaction) term in the exponential in (\ref{3.5}) and
add a corresponding expression without spin interaction for the ghosts.
Thus one arrives at
\bear\label{3.14}
V&=&\int^\infty_0\frac{dT}{T}T^{-3/2}
e^{-m^2T}\int D\bar y\exp\left\{-\int^1_0d
\bar\tau\frac{\dot{\bar y}^2}{4}\right\}
\times\nonumber\\
&&\left[-\frac{9}{2}\exp\left\{\frac{-\sigma\bar AT^3}
{(\tilde c \frac{a^2}{\bar A})^2+T^2}+\frac{4}{9}<g^2_3F^2>T^2
\int^1_0d\bar\tau d\bar\tau'D\left(
\frac{(\bar y-\bar y')^2T}{a^2}\right)\right\}\right.\nonumber\\
&&\left.+\frac{6}{2}\exp\left
\{\frac{-\sigma\bar AT^3}{(\tilde c\frac
{a^2}{\bar A})^2+T^2}\right\}\right]\ear
This is an interesting structure but it seems to be impossible
to treat the functional integral analytically. For our numerical 
estimates we replace
the $\bar y$-integral by the integrand at $\bar A=<\bar A>$.
We will come back to this point in the
discussion in section 4.

\begin{figure}[t] 
\begin{picture}(200,210)
\put(-180,10){\epsfxsize8cm \epsffile{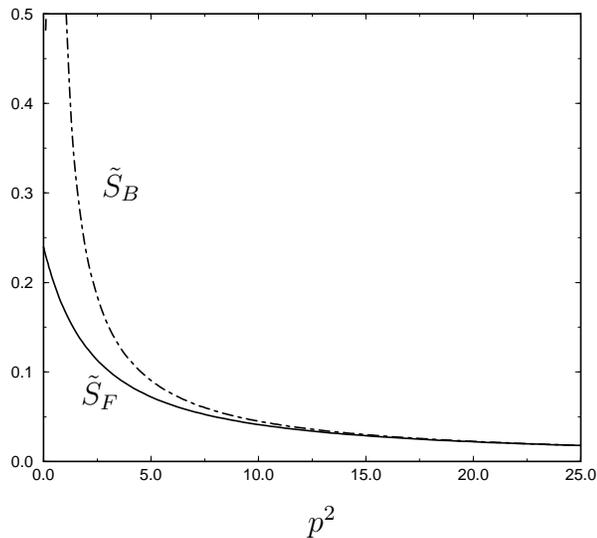}}
\put(212,20){ $p^2$}
\put(138,147){$\tilde{S}_B$}
\put(130,68){$\tilde{S}_F$}
\end{picture} 
\caption{The Euler-Heisenberg $\tilde{S}_B(p^2,m^2)$ compared to 
$\tilde{S}_F(p^2,m^2)$ for $m^2=0$ (all in $g_3^2$ units)}
\label{trot} 
\end{figure}

If the spin-spin interaction has a shorter range than the confinement
effects one could work with the integrand of (\ref{3.10'}) as a rough 
estimate for momenta $p\gsim 1/a$. At even shorter
ranges (large $p\gg 1/a$) one might suspect that (\ref{3.10'})
is not appropriate any longer: The correlation via the form factor $D$
in (\ref{1.3}) can be neglected and we have the case of a constant uncorrelated
background. If we assume this background to be
pseudoabelian, it can be accounted for by an Euler-Heisenberg
type formula. In the case of $d=3$ we
evaluate \cite{15,13c} the gauge field and ghost
contributions as\footnote{Renormalization in the first of refs.
\cite{15} after
eliminating the unstable mode is not correct.}
\bear\label{3.15}
V^{\rm EH}&=&-\frac{g_3|B|}{8\pi^{3/2}}\int^\infty_0dT T^{-3/2}
e^{-m^2T}{\Big\{}\frac{1}{\sinh(g_3|B|T)}
-\frac{1}{g_3|B|T}+\nonumber\\
&&4\sinh(g_3|B|T){\Big \}}-\frac{9-6}{2}
\frac{1}{8\pi^{3/2}}\int^\infty_0\frac{dT}{T}
T^{-3/2}e^{-m^2T}\ear
After renormalization, the last term is just the
well-known cubic term $-\frac{1}{4\pi}(m^2)^{3/2}\sim\varphi^3$.
The first integral in (\ref{3.15}) is UV
and IR convergent. For $m^2<g_3|B|$
we have an imaginary contribution due to a cut
in $(m^2-g_3|B|)^{1/2}$ signalizing instability
of the background field. This seems to be similar to the case  
$\tilde{\cS}_F(p^2,m^2)>m^2$ in eq.
(\ref{3.10}).  As is clear from the above,
(\ref{3.15}) is only justified below some IR cut-off
$T_0\sim a^2/\nu^2$, i.~e.~for momenta $p>p_0=\nu/a$,
where $\nu={\cal O}(1)$.
One can again replace the $T$ integration by a $p$
integration using the $T^{-3/2}$ substitution trick. (\ref{3.15})
can be brought to the form (\ref{3.10'}) with a
new polarization function
$\tilde {\cal S}_B(p^2,m^2)$ instead of $\tilde {\cS}_F$.
$\tilde {\cS}_B$
is compared to $\tilde {\cS}_F$ in fig. \ref{trot}. For $p^2\stackrel
{\scriptstyle>}{\sim}2/a$ indeed both quantitites
converge to each other. This is as expected since for these momenta  only
the lowest order in $\tilde {\cS}_F\sim F^2$ contributes
effectively and the nonlocality of the condensate does not play
a role any more.

\section{Evaluation and Discussion}
\setcounter{equation}{0}
As outlined in the previous chapter, even the 1-loop effective
action generated by gauge bosons and ghosts propagating in
the nonperturbative
vacuum which is modelled by a correlated gauge field and a constant
Higgs field background is a complicated expression which cannot
be evaluated analytically without drastic approximations. From
our experience with perturbative calculations
of the thermal electroweak effective potential we expect that
also (perturbative) 2-loop contributions are necessary in order
to arrive at  reliable (gauge-fixing and renormalization
point insensitive)
results. It also might turn out necessary to include higher
cumulants in a description of the long-wave length modes
by means of the stochastic vacuum model. 

If we had such an
improved effective potential at our disposal, we then
could actually determine the ``local'' gauge
condensate $<g^2F^2>$ as a function of $\varphi$
by minimizing the effective potential
in $F^2$ and in $\varphi^2\propto m^2$. Without
a detailed evaluation of the pertinent world-line path integrals
(using lattice methods, for instance)
we only can advocate a qualitative
picture: Due to spin forces the state with
$<F^2>=0$ is unstable while the true vacuum is
stabilized at some value  $<F^2>\not=0$ due to 
confinement forces.

In the hot electroweak phase $(\varphi^2\sim0)$, we
learned from lattice studies that the confining string
tension is approximately that of pure $SU(2)$-YM
theory. In the stochastic vacuum model there is the
relation (\ref{3.6})
between $<g^2_3F^2>$ and the string tension $\sigma$. Thus we may
deduce the value of the condensate from the lattice value of
$\sigma$ rather than by minimizing the effective potential.

For $m^2\not=0$ the destabilizing effect of the 1-loop $F^2$ term
diminishes (see eq. (\ref{2.16})) and at some $m^2_0$ the
$<F^2>=0$ vacuum stabilizes. Clearly $<F^2>$ depends on $m^2$.
Again $m^2_0$ could in principle be read off from a balance
between tree and loop $F^2$ terms, but as we said before, in praxi
a 1-loop calculation is not very reliable. We thus will parametrize
the $m^2$-dependence of $<F^2>$ by
\be\label{4.1}<F^2>\hspace{-0.1cm}(m^2)=<F^2>\hspace{-0.1cm}(0)h(m^2)\ee
where the function $h$ satisfies
$h(0)=1,\quad h(m^2_0)=0, \quad h'(m_0^2)=0 $.
We choose it in the form
\be\label{4.2}
h(m^2)=\cos^2(\frac{\pi}{2}\frac{m}{m_0})\ee
for $m\leq m_0$, 
and fix\footnote{In all the numerical evaluations we present in 
section 4 and appendix A we used the form 
factor $D^{(1)}$ of eq. (\ref{2.10a}).}
$<g^2_3F^2>(0)=\frac{24}{\pi}a^{-2}\sigma_{\rm fund}(m^2=0)$. The numerical
values $\sigma_{\rm fund}(m^2=0)\sim 0.13g^4_3,\quad 1/a\sim
0.73g^2_3$ are known from lattice studies \cite{13}. For
the new parameter $m_0$
in (\ref{4.2}) we choose $m_0=4a^{-1}$ so that $F^2(m^2)$ has
a mean ``decay width'' $2/a$.

In general the correlation length $a$ also
will depend on $\varphi^2\propto m^2$. For large $m^2$ the perturbative
part of the form factor $D$ (with scale $1/m$) which is not
included in our analysis (because it is represented by the
usual Feynman graphs) dominates. A dependence $a(m^2)$ should come
out from a self-consistency equation for the 
stochastic vacuum \cite{23}, but this is beyond the scope of
this paper. In the practical evaluation of our formulas $a$
is always accompanied by a factor
$<g^2_3F^2>\hspace{-0.1cm}(m^2)$ which we had to parametrize anyway in lack
of a fully dynamical treatment. Thus we keep the correlation length
constant.\footnote{A simple substitution $\frac{1}{a^2(m^2)}=
\frac{1}{a^2(0)}+m^2$ together with (\ref{4.1})/(\ref{4.2})
turned out to give a very rapid fall-off of
$m^2_{\rm conf}(p^2,m^2)$ and $S_F(p^2,m^2)$ in  $m^2$.}

\begin{figure}[t] 
\begin{picture}(200,210)
\put(-240,10){\epsfxsize7cm \epsffile{fig5a.eps}}
\put(-5,10){\epsfxsize7cm \epsffile{fig5b.eps}}
\put(325,20){ $m^2[g_3^4]$}
\put(175,170){a)}
\put(410,170){b)}
\put(90,20){ $p^2[g_3^4]$}
\put(32,170){$p^2$}
\put(277,120){$m^2_{\rm conf}$}
\put(266,66){$\tilde{S}_F$}
\put(33,55){$m^2_{\rm conf}$}
\put(98,80){$\tilde{S}_F$}
\end{picture} 
\caption{$m_{\rm conf}^2(p^2,\bar{A}=2.5,m^2)$ and $\tilde{S}_F(p^2,m^2)$ 
plotted a) for $m^2=0$ and b) $p^2=0$; we used form factor $D^{(1)}$.}
\label{SFM} 
\end{figure}

As outlined in section 3 and appendix A the gauge boson vacuum
polarization $\Sigma$ contains a destabilizing spin force
effect ${\cS}_F$. In the case of a nearest neighbour interaction
in a sufficiently extended loop we obtain a negative IR magnetic
(mass)$^2$. $\Sigma$ also contains a confining part. The latter we
have summarized into a (modified) area law exponential in
the loop path integral (\ref{3.12}). It leads to a positive
IR magnetic (mass)$^2$, $m^2_{\rm conf}(p^2)$. Such a positive
(mass)$^2$ in the IR region acts as an IR cutoff. Indeed, the soft
gauge quanta are already taken into account by the
stochastic vacuum background. They should  not
circulate in the gauge boson loop of the effective action.

There seems to be a paradoxon: On one hand, in order to
destabilize $<F^2>=0$, the spin forces have to dominate 
the $F^2$ radiative correction calculated in section 2. On the
other hand confinement forces should stabilize $F^2$ at some
value $<F^2>\not=0$ and there these forces should dominate,
i.e. the singularity in $p^2$ shifted towards positive $p^2$
by the $\tilde {\cS}_F$ term in (\ref{3.10}), (\ref{3.10'})
should not be reached because now the positive $m^2_{\rm conf}$
dominates. However, this is not a contradiction if
$\tilde {\cS}_F(p^2)$ and $m^2_{\rm conf}(p^2)$ have different $p^2$
dependences and dominate at intermediate and small
$(p\stackrel{\scriptstyle<}{\sim}\frac{1}{a})$
loop momenta, respectively. For suitable values of
$<\bar A>$, this can be indeed the case
as we see from fig. \ref{SFM}. Here we have chosen $<\bar A>
=2.5,\ \tilde c=2$ in (\ref{3.11})
and $m^2=0$.

It is important to note that also $\tilde {\cS}_F$ is
changed (diminished!)
by the effective IR cut-off through $m^2_{\rm conf}((p\pm q)^2)$
entering (\ref{3.9}). If $m^2_{\rm conf}$ is too
large, $\tilde {\cS}_F$
is strongly suppressed and never dominates -- this would exclude
the destabilization of $<F^2>=0$! 

The size of
the $m^2_{\rm conf}(p^2=0)$ chosen above is of the same order as the
gap mass $m_{\rm NJL}$ of refs. \cite{21} who give a pole mass
$m_{\rm NJL}=0.28 g^2_3$ for
``the magnetic mass'' in a 1-loop Nambu-Jona-Lasinio-type
equation for the gluon propagator. It is also in the range of
the Landau gauge gluon propagator mass, $m_L=0.35g_3^2$ in the 
lattice evaluation
of \cite{22} and of the magnetic mass obtained from gap equations
for pure YM theory \cite{nair},\cite{corn}. Extrapolating our 
$m^2_{\rm conf}(p^2)$ to the
``pole-mass'' (at negative $p^2=-m^2_{\rm conf}(p^2)$)
would increase its value according to fig. \ref{SFM}.

We can bring together the negative $-\tilde S_F(p^2)$ and the
positive $m^2_{\rm conf}(p^2)$ in one expression in
order to obtain a rough
approximation to the gauge particle contribution to the 1-loop
effective potential. We also add the ghost part which  contains
the $m^2_{\rm conf}(p^2)$-modification only:
\bear\label{4.4}
V_1^g&=&\frac{9}{2}\int^\infty_0\frac{dpp^2}{2\pi^2}\left\{
2\ln[p^2+m^2+m^2_{\rm conf}(p^2,m^2)-\tilde {\cS}_F(p^2,m^2)]
\right.\nonumber\\
&&-\ln[p^2+m^2+m^2_{\rm conf}(p^2,m^2)]+\frac{\tilde {\cS}_F
(p^2,m^2)}{p^2+m^2+m^2_{\rm conf}(p^2,m^2)}
-\frac{m^2}{p^2}-\ln p^2{\Big\}}\nonumber\\
&&-\frac{6}{2}\int^\infty_0\frac{dpp^2}{2\pi^2}\left\{\ln
[p^2+m^2+m^2_{\rm conf}(p^2,m^2)]-\frac{m^2}{p^2}-\ln p^2\right\}
\ear
with the modified $\tilde {\cS}_F$ of (\ref{3.9})
\bear\label{4.5}
\tilde {\cS}_F(p^2,m^2)&=&\frac{<g^2_3FF>\hspace{-0.1cm}(m^2)}{9\pi^2}
\int^\infty_0dqq^2\frac{1}{pq}
\nonumber\\
&&\ln\left(\frac{(p+q)^2+m^2+m^2_{\rm conf}((p+q)^2,m^2)}
{(p-q)^2+m^2+m^2_{\rm conf}((p-q)^2,m^2)}\right) \tilde D(q^2)\ear
As we argued before (fig. \ref{trot}) also the
Euler-Heisenberg region $p\gsim\frac{3}{a}$ is properly included.

A similar 1-loop potential due to scalars (Higgs plus Goldstone
bosons) in the loop has the form
\bear\label{4.5'}
&&V_1^H=\frac{1}{2}\int\frac{dp p^2}{2\pi^2}\left\{\left[
\ln(p^2+m^2_3+3\lambda\varphi^2+m^2_{\rm conf(fund)}(p^2,m^2))
-\ln p^2-\frac{m^2_3+3\lambda\varphi^2}{p^2}\right]\right.\nonumber\\
&&\left.+3\left[\ln(p^2+m^2_3+\lambda\varphi^2+m^2+m^2_{\rm conf(fund)}(p^2,
m^2))-\ln p^2-\frac{m^2_3+\lambda\varphi^2+m^2}{p^2}\right]\right\}
\ear
$m^2_{{\rm conf
(fund)}}$ is related to $m^2_{\rm conf}$ via the relation
between the string tension in the adjoint and fundamental
representation (\ref{3.6}) as
\be\label{4.5''}
m^2_{\rm conf(fund)}(p^2,m^2)=\frac{3}{8}m^2_{\rm conf}(p^2,m^2)\ee
In the evaluation of (\ref{4.5'}) we vary $m^2_3$ starting
from positive values (high temperature) and avoid the
unstable part with negative $m^2_3$ where a treatment in the framework
of coarse grained actions seems to be indispensible.

\begin{figure}[t] 
\begin{picture}(200,210)
\put(-180,10){\epsfxsize8cm \epsffile{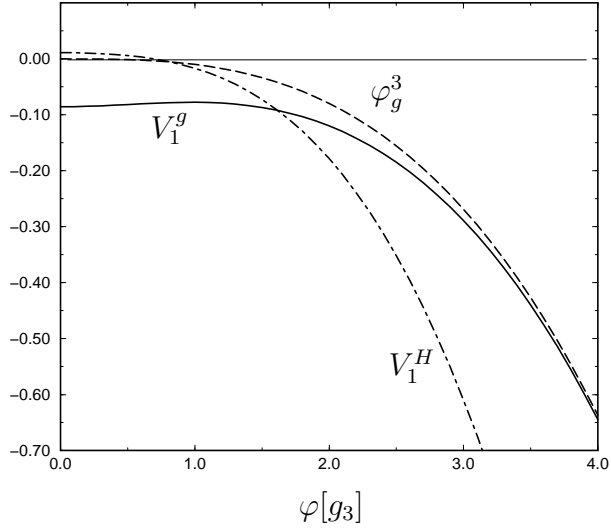}}
\put(202,22){ $\varphi[g_3]$}
\put(240,75){$V_1^H$}
\put(150,165){$V_1^g$}
\put(233,179){$\varphi^3_g$}
\end{picture} 
\caption{$V_1^g(\varphi)$, $V_1^H(\varphi)$ compared to the perturbative
$\varphi^3_g$ term}
\label{V1} 
\end{figure}

The 1-loop potentials (\ref{4.4}) and (\ref{4.5'}) are
the main tool for our
investigation of the phase
structure of the hot electroweak theory.
They are plotted in fig. 6 for $\lambda=0.11g_3^2$ and $m_3^2=-0.0041g_3^4$ 
together with
the perturbative $\varphi^3$-term. They have to be added to the
tree Higgs potential
\be\label{4.6}
V(\varphi_{\rm DL})=\frac{1}{2}y\varphi^2_{\rm DL}+\frac{1}{4}x
\varphi^4_{\rm DL}+V^{\rm tree}_{FF}+(V_1^g+V_1^H)/(g^2_3)^3\ee
written with the dimensionless quantities (\ref{1.2}) and the
dimensionless field $\varphi^2_{\rm DL}=\varphi^2/g^2_3\ (m^2=g^2_3
\varphi^2/4=g^4_3\varphi^2_{\rm DL}/4)$. $V_{FF}^{\rm tree}$ of
eq. (\ref{2.22}) turns out to be small numerically. 

\begin{figure}[t] 
\begin{picture}(200,210)
\put(-180,10){\epsfxsize8cm \epsffile{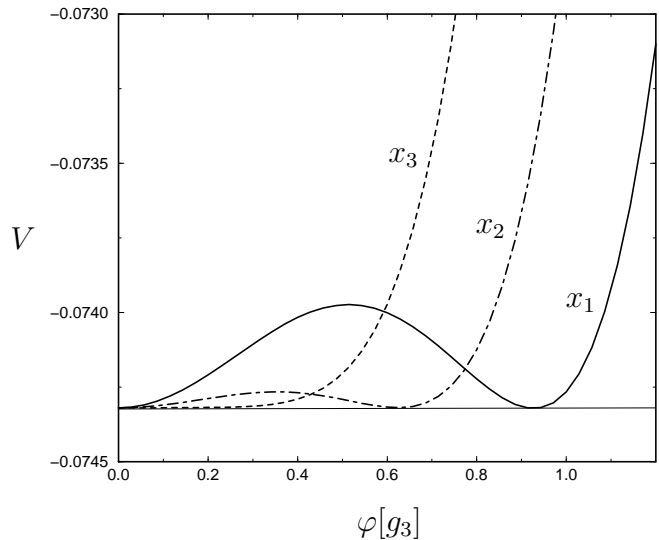}}
\put(202,20){ $\varphi[g_3]$}
\put(75,127){$V$}
\put(218,160){$x_3$}
\put(251,133){$x_2$}
\put(285,105){$x_1$}
\end{picture} 
\caption{Fading away of the first order phase transition
with increasing $x=\frac{\lambda}{g_3^2}$, where $x_1=0.06$,
$x_2=0.08$ and $x_3=0.11$}
\label{Vtot} 
\end{figure}

As we can see in fig. 7, starting at small $x$
the potential (\ref{4.6}) signals the usual first-order phase transition,
i.e. two degenerate minima at the appropriate $y$ (temperature $T_c$)
and in between a bump with a maximum. Increasing $x$ the bump
between the minima and the ``size of the broken phase'' $\varphi_{min}$
becomes smaller. In the case $\bar A=2.5$, $\tilde c=2$ and $m_0=4a^{-1}$
the bump vanishes at
$x\sim 0.11$. This is  in agreement with lattice calculations: we
have a crossover. For this result the $\tilde {\cS}_F$ spin part
is very important.\footnote{For the parameter set used in fig.~\ref{Vtot} 
$m^2_{\rm conf}-\tilde S_F$ is of the order of $m^2_{\rm NJL}$ obtained in ref.
\cite{21}. The Higgs part of the potential just contains $\frac{3}{8}m_{\rm conf}^2$
which is also of that order whereas the ghost part has the full $m^2_{\rm conf}$.} 

In the work of Buchm\"uller and Philipsen  \cite{21}
the crossover point was determined by comparing the ``magnetic mass''
$m^2_{\rm NJL}$ of the gauge bosons in the hot phase with
the (perturbative) gauge boson mass in the broken
phase in the unitary gauge. 
Since our ``magnetic mass'' is momentum dependent this is not 
a clear prescription in our case. 
Trying a similar reasoning
we could compare $m^2_{\rm conf}(p^2=0,\ m^2=0)-\Sigma
(p^2=0, m^2=0)$ with
the perturbative gauge boson mass in the ``broken'' phase.

At the endpoint of the first-order phase transition line 
in the $x-y$ plane (``crossover point'')
 we expect a second-order phase transition
and thus conformal symmetry. The latter leads to the
requirement that
the dimensionful parameters $m^2_{3_{\rm eff}}$ and $\lambda_{\rm eff}$
multiplying the $\phi^2$- and $\phi^4$- terms
of the effective 3-dimensional scalar action should be
zero at the crossover point. At the present level of accuracy we can be 
gentle about anomalous 
dimensions which may indeed be rather small \cite{RTW}.
Taylor expanding the potential $V$ of eq. (\ref{4.6})
allows us to read off the coupling $\lambda_{\rm eff}$ and with the parameters
chosen in fig. 7 we obtain
\be\label{4.5'''}
x_{\rm cro}=0.11 \ee 
It is an interesting question whether the second-order phase transition
is in the universality class of scalar $\phi^4$ theory and the 
Ising model \cite{RTW}.\footnote{This was reported recently as
a lattice result \cite{klr}.}   
Of course our approach is not quite adequate to
give a satisfactory answer to this. 

The table below contains the contribution of $V_1^g$ and $V_1^H$
to the 1-loop $\varphi^4$ term and the effective coupling at the
$x$ values of fig.~\ref{Vtot}. We also give the surface tension in the 
thin wall approximation 
$\sigma_{\rm s}=\int_{\varphi_S}^{\varphi_A}\sqrt{2V(\varphi)}d\varphi$
for the full and the perturbative potential. We get the qualitative
features of the comparison of perturbative and lattice results for
$\sigma_{\rm s}$ \cite{5}.  

\vspace{0.5cm}
\centerline{
\begin{tabular}{|r||r|r|r|r|r|} \hline
$x$ & $\delta x^g$& $\delta x^H$ & $x_{\rm eff}$ & $\sigma_{\rm s}$ & 
$\sigma_{\rm s,per}$ \\ \hline \hline
0.06 & -0.063 & -0.039 & -0.041 & 0.016 & 0.013 \\ \hline
0.08 & -0.063 & -0.043 & -0.026 & 0.004 & 0.007 \\ \hline
0.11 & -0.063 & -0.051 & 0 & 0 & 0.004 \\ \hline
\end{tabular} }
\vspace{0.5cm}

Our most important calculation of the $F^2$ term in chapter 2 only 
contained the correlation length $a$ which we took from lattice data;
there is also some uncertainty in this term because of its gauge
dependence which should be reduced by a 2-loop calculation.
The construction of an effective potential in 1-loop order contained
many more parameters: $<F^2>(0)$ and the function $h(m^2)$ 
containing $m_0^2$ in (\ref{4.1}) could in principle be obtained 
by studying the nontrivial minimum of the effective potential
in $F^2$; $\tilde{c}$ and $n=3$ in (\ref{3.12}) are related to the
modification of the area law at small areas. $<\bar{A}>$ could be
determined in a numerical study of the path integral (\ref{3.14}).
Given our ignorance about the last three parameters
$\tilde{c}$, $n$ and $<\bar{A}>$ we just fitted a potential
in order to reproduce the crossover at  the $x$-value given in
lattice calculations. We then did not consistently determine
$<F^2>(0)$ and $h(m^2)$ from the potential but fixed
the former by lattice data on the string tension and used 
a plausible value for $m_0^2$. This point requires more exhaustive
numerical studies, not to speak about a 2-loop calculation again.

We have described the IR effects of the gauge interactions by
use of the stochastic vacuum correlation (\ref{1.3}). This is quite
different from a renormalization group treatment like in ref.
\cite{2a} where differential renormalization group equations
(at least in principle) lead to an effective IR action starting
from the usual tree action in the UV. 
The stochastic vacuum prescription is a pure IR concept 
and - in a sense - complementary to the renormalization
group approach. The latter is applied most easily if the
relevant momentum scale is not too far below the UV
cutoff where the initial condition (classical action) is specified.
In this situation simple truncations of the space of actions 
are quite successful since the effective (average) action
does not deviate very strongly from the classical one.
Once the variable IR cutoff enters the deep IR-regime
a more exhaustive parametrization of the space of actions
becomes indispensable. For pure Yang-Mills theory, a
first step in this direction has been done in ref. \cite{gluco},
but a complete ab initio treatment of the IR physics is
necessarily quite involved and not available yet.

The stochastic vacuum model, on the other hand, makes
a phenomenological ansatz for the low momentum modes
of the gauge field. In principle this effective IR dynamics
should follow from the renormalization group analysis
but clearly the actual proof is a formidable task. Accepting
the stochastic vacuum as a starting point one is in the
opposite situation as in the renormalization group 
framework where one is given an effective average action
at a certain scale $k$. In the former case, only the modes with
very small momenta are described by the correlators used 
and all other modes with {\it larger} momenta, have to be
dealt with (``integrated out'') explicitly, e.~g.~by perturbation
theory in the background of the low-momentum modes.
This is what we did in sec. 2, for instance, when we
stochastically averaged the one-loop effective action.
Conversely, in the renormalization group case the
modes with momenta {\it smaller} than $k$ are the ones which
still have to be integrated out.

It would also be very interesting to see if the stochastic 
vacuum model can be derived from a
self-consistent set of Dyson-Schwinger equations as
proposed in ref. \cite{23}. 

Our approach is also quite different
from the Nambu-Jona-Lasinio-type equations 
for the ``magnetic mass'' which are proposed in \cite{21}. 
In the latter approach, this magnetic
mass seems to have a similar meaning as the confinement
mass $m^2_{\rm conf}$
introduced in our discussion. We argued that this is an effective
(mass)$^2$, which is only important for very small momenta and we
have an additional negative $\mbox{(mass)}^2$, $-\tilde {\cS}_F$, 
related to the gauge boson spin-spin interaction
which is in a delicate balance with the
confinement mass. 
This avoids the appearence of a tachyonic pole in the gluon
propagator related to the well known IR instability of YM theory.
A higher (two) loop calculation in the nonperturbative gauge
field background would be desirable (though very demanding)
both in order to reduce the gauge dependence and in order to
treat the IR instability properly. Indeed this also seems to be
an important point in the discussion of gauge invariant gap
equations \cite{nair,21} in a recent paper of Cornwall \cite{corn}.
It would be interesting to include a gauge
background in the gap equations of ref. \cite{21}. 

In our approach
the nonperturbative dynamics mainly enters through the correlated
gauge field background like in pure gauge theory but with a further
Higgs field parameter $m^2\propto\varphi^2$. We have been using 
lattice results for the values of $a$ and $\sigma_{\rm fund}(m=0)$. 
It might be possible to explore the $m^2$ dependence of these 
quantities on the lattice.

In finite temperature QCD (and in ordinary QCD) we have no Higgs
expectation value $\varphi^2$ at our disposal which could act as 
an IR regulator and the 4-dimensional gauge coupling is big. 
Thus the status
of naive perturbation theory is even much worse \cite{Kaj}. Still
an IR-improved perturbative picture using a correlated gauge field
background might be possible \cite{Simonov} and the methods 
introduced in the present paper may be still relevant. This 
deserves further investigation.

\section*{Acknowledgement}
We would like to thank B. Bergerhoff, D. B\"odeker, H. G. Dosch, 
M. Ilgenfritz, M. Laine, O. Philipsen, Y. Simonov, and C. Wetterich 
for useful discussions and J. Kripfganz for collaboration in an early
stage. This work was supported in part by the
TMR network {\it Finite Temperature Phase Transitions in Particle 
Physics}, EU contract no. ERBFMRXCT97-0122.

\section*{Appendix A}
\renewcommand{\theequation}{A.\arabic{equation}}
\setcounter{equation}{0}

In this appendix we calculate the stochastic average of the gauge
boson propagator $G=K^{-1}$ with the kinetic operator $K^{ab}
_{\mu\nu}$ given in eq. (\ref{2.12}).  We shall determine $G$ to first
order in the condensate $<g^2F^2>$.

To start with, we decompose $K_{\mu\nu}$ according to
\be\label{A.1}
K_{\mu\nu}=K^{(0)}_{\mu\nu}+K^{(1)}_{\mu\nu}+K^{(2)}_{\mu\nu}\ee
where the superscripts denote the powers of $F_{\mu\nu}$ which
are contained in the respective terms once the gauge field
\be\label{A.2}
A_\mu(x)=\int^1_0d\eta\ \eta x_\nu F_{\nu\mu}(\eta x)\ee
is inserted. Thus we employ the Fock-Schwinger gauge
centered at $x_\mu=0$ 
\bear\label{A.3}
&&K_{\mu\nu}^{(0)}=(\hat p^2+m^2)\delta_{\mu\nu}\nonumber\\
&&K_{\mu\nu}^{(1)}=-g(\hat p_\alpha A_\alpha +A_\alpha\hat p_\alpha)
\delta_{\mu\nu}+2ig F_{\mu\nu}\nonumber\\
&&K_{\mu\nu}^{(2)}=g^2A_\alpha A_\alpha \delta_{\mu\nu}\ear
with $\hat p_\mu\equiv -i\partial_\mu$ and with
$A_\mu$ and $F_{\mu\nu}$ considered as matrices in the adjoint
representation of the gauge group.

To order $F^2$, the propagator then reads
\bear\label{A.4}
G&=&G^{(0)}-G^{(0)}K^{(1)} G^{(0)}-G^{(0)}K^{(2)}G^{(0)}\nonumber\\
&&+G^{(0)}K^{(1)}G^{(0)}K^{(1)}G^{(0)}+0(F^3)\ear
where $G^{(0)}=[K^{(0)}]^{-1}$ is the free propagator.

Using (\ref{1.3}) together with $\ll1\gg=1$ and $\ll F_{\mu\nu}\gg=0$ the
stochastic average $\bar G\equiv \ll G\gg$ is seen to be
\be\label{A.5}
\bar G=G^{(0)}+G^{(0)}\Sigma G^{(0)}+0(F^3)\ee
i.e.,
\be\label{A.6}
\bar G^{-1}=G^{(0)-1}-\Sigma+0(F^3),\ee
with the ``mass operator''
\be\label{A.7}
\Sigma\equiv \ll K^{(1)}G^{(0)}K^{(1)}\gg-\ll K^{(2)}\gg\ee
The quadratic averages in (\ref{A.7}) have to be performed using the
correlator (\ref{1.3}). For simplicity we restrict ourselves to the case
$\kappa=1$. Furthermore, we choose the reference point
$x_0=0$, which coincides with the point at which the Fock-Schwinger
gauge is centered. As a consequence, the path-ordered
exponential in eq. (\ref{1.4}) equals unity and it
follows that
\bear\label{A.8}
&&\ll g^2F^a_{\mu\nu}(x)F^b_{\rho\sigma}(y)\gg
=\omega\delta^{ab}\nonumber\\
&&\qquad \times[\delta_{\mu\rho}\delta_{\nu\sigma}-\delta_{\mu\sigma}
\delta_{\nu\rho}]D((x-y)^2/a^2)\ear
with
\be\label{A.9}
\omega\equiv\frac{<g^2F^2>}{d(d-1)(N_c^2-1)}\ee
It is convenient to write
\be\label{A.10}
\Sigma=\Sigma_1+\Sigma_2\ee
\be\label{A.11}
\Sigma_1\equiv\ll Q\gg,\quad Q\equiv K^{(1)}
G^{(0)}K^{(1)}\ee
\be\label{A.12}
\Sigma_2\equiv-\ll K^{(2)}\gg\ee
and to further decompose
\be\label{A.13}
Q=Q_{AA}+Q_{FF}+Q_{AF}\ee
with
\be\label{A.14}
{}[Q_{AA}]_{\mu\nu}=g^2(\hat p_\alpha A_\alpha+A_\alpha\hat p_\alpha)
G^{(0)}_{\mu\nu}(\hat p)(\hat p_\beta A_\beta+A_\beta\hat p_\beta)\ee
\be\label{A.15}
{}[Q_{FF}]_{\mu\nu}=(2ig)^2F_{\mu\alpha}G^{(0)}_{\alpha\beta}(\hat p)F_
{\beta\nu}\ee
\bear\label{A.16}
{}[Q_{AF}]_{\mu\nu}&=&-2ig^2\Bigl\{(\hat p_\alpha A_\alpha+A_\alpha
\hat p_\alpha)G^{(0)}_{\mu\beta}(\hat p)F_{\beta\nu}\nonumber\\
&&+F_{\mu\alpha}G^{(0)}_{\alpha\nu}(\hat p)(\hat p_\beta
A_\beta+A_\beta\hat p_\beta)\Bigr\}\ear
Working in a position-space representation, quantities such as
$\Sigma_1$ or $Q$ are kernels with respect to the
space-time coordinates and matrices with respect to Lorentz
and adjoint group indices. Before we can evaluate the average
\be\label{A.17}
\Sigma_1(x_2,x_1)^{ab}_{\mu\nu}=\ll Q(x_2,x_1)^{ab}_{\mu\nu}\gg\ee
we have to determine the position-space matrix elements
\be\label{A.18}
Q(x_2,x_1)^{ab}_{\mu\nu}=<x_2|Q^{ab}_{\mu\nu}(\hat x,\hat p)
|x_1>\ee
for a fixed background field. For the contribution coming from
$Q_{AA}$, say, one finds
\bear\label{A.19}
&&Q_{AA}(x_2,x_1)^{ab}_{\mu\nu}=-g^2(T^cT^d)^{ab}\delta_{\mu\nu}
\nonumber\\
&&{}\qquad\times\left[4A^c_\rho(x_2)A^d_\sigma(x_1)\partial_\rho
\partial_\sigma G^{(0)}(x_2-x_1)\right.\nonumber\\
&&\hspace{0.5cm}{}\qquad-2A^c_\rho(x_2)\partial A^d(x_1)\partial_\rho G^{(0)}
(x_2-x_1)\nonumber\\
&&\hspace{0.5cm}{}\qquad+2\partial A^c(x_2)A^d_\sigma(x_1)\partial_\sigma
G^{(0)}(x_2-x_1)\nonumber\\
&&\hspace{0.5cm}{}\qquad\left.-\partial A^c(x_2)\partial A^d(x_1)
G^{(0)}(x_2-x_1)\right]\ear
with $\partial A^a\equiv\partial_\mu A^a_\mu$ and $<x_2|G^{(0)}_{\mu\nu}
|x_1>\equiv \delta_{\mu\nu} G^{(0)}(x_2-x_1)$.
The next step is to express all $A_\mu$'s in terms of
$F_{\mu\nu}$'s by using (\ref{A.2}), and to apply (\ref{A.8})
to the resulting $\ll FF\gg$ correlators. In particular one
needs
\bear\label{A.20}
&&\ll g^2A^c_\mu(x_2)A^d_\nu(x_1)\gg=\omega\delta^{cd}[(x_1\cdot
x_2)\delta_{\mu\nu}-x_{1\mu}x_{2\nu}]\nonumber\\
&&\qquad\times\int^1_0d\eta_1\int^1_0d\eta_2\eta_1\eta_2 D(\delta^2/a^2)
\nonumber\\
&&\ll g^2A^c_\mu(x_2)\partial A^d(x_1)\gg=-2\omega\delta^{cd}a^{-2}
[(x_1\cdot x_2)x_{2\mu}-(x_2)^2x_{1\mu}]\nonumber\\
&&\qquad\times\int^1_0d\eta_1\int^1_0d\eta_2(\eta_1\eta_2)^2D'(\delta^2/
a^2)\nonumber\\
&&\ll g^2\partial A^c(x_2)\partial A^d(x_1)\gg=-2\omega\delta^{cd}a^{-2}
\nonumber\\
&&\qquad\times\left\{(d-1)(x_1\cdot x_2)\int^1_0d\eta_1\int^1_0d\eta_2
(\eta_1\eta_2)^2D'(\delta^2/a^2)\right.\nonumber\\
&&\hspace{1.4cm}\left.+2a^{-2} v_{12}\int^1_0d\eta_1\int^1_0d\eta_2(\eta_1\eta_2)^3
D''(\delta^2/a^2)\right\}\ear
with the abbreviations
\be\label{A.21}
\delta_\mu\equiv \eta_2(x_2)_\mu-\eta_1(x_1)_\mu\ee
and
\be\label{A.22}
v_{12}\equiv (x_1)^2(x_2)^2-(x_1\cdot x_2)^2\ee
The derivatives acting on $G^{(0)}$ in eq. (\ref{A.19})
can be easily removed by noting that the free propagator
has the structure
\be\label{A.23}
G^{(0)}(x_2-x_1)=H\left(\frac{1}{2}(x_2-x_1)^2\right)\ee
for some function $H$. In the massless case, for instance,
\be\label{A.23a}
H(\nu)=\frac{1}{2}(2\pi)^{-d/2}\Gamma(d/2-1)\ \nu^{-d/2+1},
\quad d>2\ee
After some calculation this leads to
\bear\label{A.24}
&&\ll Q_{AA}(x_2,x_1)^{ab}_{\mu\nu}\gg=-2\omega N_c\delta^{ab}
\delta_{\mu\nu}\nonumber\\
&&\qquad\times\left\{2[(d-1)(x_1\cdot x_2)H'+v_{12}H'']\int^1_0
d\eta_1d\eta_2\eta_1\eta_2D(\delta^2/a^2)\right.\nonumber\\
&&\hspace{0.6cm}\qquad+4a^{-2}v_{12}H'\int^1_0d\eta_1d\eta_2(\eta_1\eta_2)^2D'(\delta
^2/a^2)\nonumber\\
&&\hspace{0.6cm}\qquad+(d-1)a^{-2}(x_1\cdot x_2)H\int^1_0d\eta_1d\eta_2
(\eta_1\eta_2)^2D'(\delta^2/a^2)\nonumber\\
&&\hspace{0.6cm}\qquad\left.+2a^{-4}v_{12}H\int^1_0d\eta_1d\eta_2(\eta_1\eta_2)^3
D''(\delta^2/a^2)\right\}\ear
Here a prime denotes the derivative with respect to the
argument, and the function $H$ and its derivatives are always
evaluated at the point $(x_2-x_1)^2/2$. Likewise one finds
for the other contributions to $\Sigma_1$:
\bear
\ll Q_{FF}(x_2,x_1)^{ab}_{\mu\nu}\gg&=&4(d-1)\omega N_c\delta^{ab}
\delta_{\mu\nu}H((x_2-x_1)^2/2)D((x_2-x_1)^2/a^2)\nonumber\\
&&\label{A.25}\\
\ll Q_{AF}(x_2,x_1)^{ab}_{\mu\nu}\gg&=&-2g^2\omega N_c\delta^{ab}
(x_{1\mu}x_{2\nu}-x_{1\nu}x_{2\mu})\nonumber\\
&&\cdot\Bigl\{H'((x_2-x_1)^2/2)\int^1_0d\eta \eta D
((x_1-\eta x_2)^2/a^2)\nonumber\\
&&+2a^{-2}H((x_2-x_1)^2/2)\int^1_0d\eta \eta^2D'((x_1-\eta x_2)^2/a^2)
\nonumber\\
&&+(x_1\leftrightarrow x_2)\Bigr\}\label{A.26}\ear
The contribution $\Sigma_2$ is a pure contact term which
vanishes for $x_1\not= x_2$:
\bear\label{A.27}
\Sigma_2(x_2,x_1)^{ab}_{\mu\nu}&=&-(d-1)\omega N_c\delta^{ab}
\delta_{\mu\nu}x^2_1\delta(x_2-x_1)\nonumber\\
&&\cdot\int^1_0d\eta_1d\eta_2\eta_1\eta_2D((\eta_1-\eta_2)^2x^2_1/a^2)
\ear

The complete ``mass operator'' $\Sigma$ is given by the sum
of the terms in eqs. (\ref{A.24}), (\ref{A.25}), (\ref{A.26}) and
(\ref{A.27}). We observe that for the gauge chosen, $\Sigma$
is not translational invariant; it depends on $x_1$ and $x_2$
separately and not only on the difference $x_1-x_2$. Generally
speaking $\Sigma$ is a function of the two translational
invariant combinations of $x_1,x_2$ and the point $x_0$ at
which the Fock-Schwinger gauge is based (cf. eq. (\ref{3.3}))
and which was chosen to act also as the reference point in the
$\ll FF\gg$-correlator, see eq. (\ref{1.3}). Without loss of
generality, one of these three points can be fixed at will. We have
exploited this freedom in order to set $x_0=0$. The general
case is recovered by substituting $x_{1,2}\to x_{1,2}-x_0$.
We see that $\Sigma$ is proportional to the unit matrix in color
space. It is not, however, a unit matrix with respect to
the Lorentz indices since $Q_{AF}$ has a nontrivial tensor structure
involving $x_1$ and $x_2$.

Next we consider $\Sigma$ with a different choice of the point
$x_0$ which, again, serves both as the reference point in
the correlator and as the base point of the Fock-Schwinger gauge.
We define $x_0$ to be in the middle of the straight line
connecting $x_1$ to $x_2$:
\be\label{A.28}
x_0=\frac{1}{2}(x_1+x_2)\ee
 For this choice of $x_0$, $\Sigma$ depends on $x_1$ and
$x_2$ only via $(x_1-x_2)^2$. One finds
\be\label{A.29}
\Sigma(x_2,x_1)^{ab}_{\mu\nu}=\delta_{\mu\nu}\delta^{ab}
{\cal S}((x_1-x_2)^2)\ee
The function ${\cal S}$ can be determined from our first
calculation by setting
\[x_1=\frac{1}{2}z,\qquad x_2=-\frac{1}{2}z\]
for some vector $z$. Then $x_0=\frac{1}{2}(x_1+x_2)=0$,
and the reference points agree. In this manner one is led
to the comparatively simple result $(z\equiv x_1-x_2)$
\bear\label{A.30}
{\cal S}(z^2)&=&\frac{N_c}{d(N_c^2-1)}<g^2F^2>\nonumber\\
&&\cdot\Bigl\{z^2H'\left(\frac{z^2}{z}
\right)\int^1_0d\eta_1d\eta_2\eta_1
\eta_2D\left((\eta_1+\eta_2)^2\frac{z^2}{4a^2}\right)
\nonumber\\
&&+\frac{z^2}{2a^2}H\left(\frac{z^2}{2}\right)
\int^1_0d\eta_1d\eta_2(\eta_1\eta_2)^2D'\left((\eta_1+\eta_2)
^2\frac{z^2}{4a^2}\right)\nonumber\\
&&+4H\left(\frac{z^2}{2}\right)
D\left(\frac{z^2}{a^2}\right)
{\Big\}}\ear
The $Q_{AF}$ contribution vanishes.

Since the operator $\Sigma$ calculated according to
the second prescription is translational invariant,
it is meaningful to Fourier-transform it with respect to $z$, and
to interpret $\tilde\Sigma(p^2=0)$ as a kind of mass shift
originating from the interaction of the gauge bosons with the
stochastic background field. If we define
\be\label{A.31}
\tilde{\cal S}(p^2)\equiv \int d^dze^{-ipz}{\cal S}(z^2)\ee
and use the Fourier transforms
\bear\label{A.32}
&&H\left(\frac{z^2}{2}\right)=\int\frac{d^dq}{(2\pi)^d}
e^{iqz}(q^2+m^2)^{-1}\nonumber\\
&&D\left(\frac{z^2}{a^2}\right)=\int\frac{d^dq}{(2\pi)^d}
e^{iqz}\tilde D(q^2)\ear
it is straightforward to arrive at
\bear\label{A.33}
&&\tilde{\cal S}(p^2)=\frac{N_c}{d(N_c^2-1)}
<g^2F^2>\int\frac{d^dq}{(2\pi)^d}\Biggl\{4\frac{\tilde D(q^2)}
{(p-q)^2+m^2}\nonumber\\
&&-2^d\int^1_0d\eta_1d\eta_2\frac{\eta_1\eta_2}{(\eta_1
+\eta_2)^d}\left[\frac{d-2}{(p-q)^2+m^2}+\frac{2m^2}{[(p-q)^2+m^2]^2}
\right]\tilde D\left(\frac{4q^2}{(\eta_1+\eta_2)^2}\right)\nonumber\\
&&-2^d\int^1_0d\eta_1d\eta_2\frac{(\eta_1\eta_2)^2}{(\eta_1
+\eta_2)^{2+d}}\left[d\tilde  
D\left(\frac{4q^2}{(\eta_1+\eta_2)^2}\right)\right.\nonumber\\
&&\left.\qquad+\frac{8q^2}{(\eta_1+\eta_2)^2}\tilde D'\left(\frac{4q^2}
{(\eta_1+\eta_2)^2}\right)\right]\frac{1}{(p-q)^2+m^2}
\Biggr\}\ear
\begin{figure}[t] 
\begin{picture}(200,210)
\put(-180,10){\epsfxsize8cm \epsffile{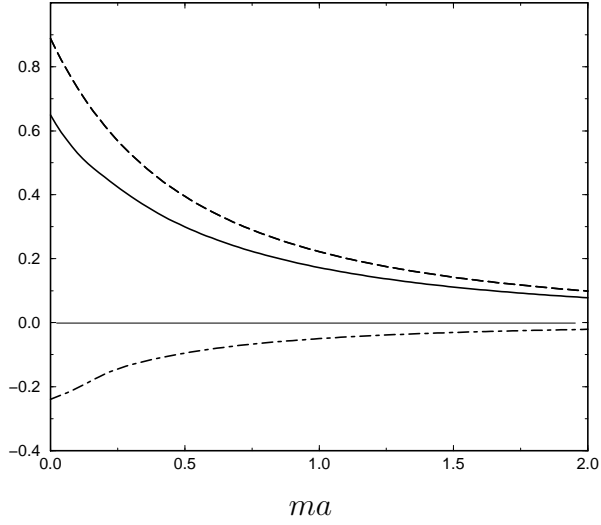}}
\put(202,22){ $ma$}
\put(240,75){$$}
\put(150,165){$$}
\put(233,179){$$}
\end{picture} 
\caption{Defining $\tilde{\cal S}(p^2,m^2)=<g_3^2F^2>a^2\cdot s(p^2,m^2)$ 
the full line is $s(p^2=0,m^2)$, the dashed and the dot-dashed lines 
are the contributions of the spin and the minimal interactions to 
$s(p^2=0,m^2)$, respectively. We used the form factor $D^{(1)}$.}
\label{sigmas} 
\end{figure}
For the remaining analysis we restrict
ourselves to the physically most
interesting case in this paper, $d=3$ and $N_c=2$. The angular 
integration in
(\ref{A.33}) can be evaluated analytically. One obtains
\bear\label{A.34}
&&\tilde{\cal S}(p^2)=\frac{2}{9\pi^2}<g^2_3F^2>\int^\infty_0dqq^2
\nonumber\\
&&\left\{\frac{1}{2pq}\tilde  
D(q^2)\ln\frac{(p+q)^2+m^2}{(p-q)^2+m^2}\right.\nonumber\\
&&-\int^1_0d\eta_1d\eta_2\frac{\eta_1\eta_2}{(\eta_1+\eta_2)^3}
\left[\frac{1}{pq}\ln\frac{(p+q)^2+m^2}{(p-q)^2+m^2}+
\frac{8m^2}{(p^2+q^2+m^2)^2-4p^2q^2}\right]
\nonumber\\
&&\hspace{2cm}\cdot\tilde D\left(\frac{4q^2}{(\eta_1+\eta_2)^2}\right)\nonumber\\
&&-\int^1_0d\eta_1d\eta_2\frac{(\eta_1\eta_2)^2}{(\eta_1+
\eta_2)^5}\frac{1}{pq}\left[3\tilde D
\left(\frac{4q^2}{(\eta_1+\eta_2)^2}\right)
+\frac{8q^2}{(\eta_1+\eta_2)^2}\tilde D'
\left(\frac{4q^2}{(\eta_1+\eta_2)^2}\right)\right]
\nonumber\\
&&\left.\hspace{2cm}\cdot\ln\frac{(p+q)^2+m^2}{(p-q)^2+m^2}\right\}
\ear
The integral representation (\ref{A.34}) is the main
result of this appendix. The remaining $q$-integration
has to be performed numerically. Results are given 
in fig. \ref{sigmas}.\footnote{For the form factor $D^{(2)}$, $s(0,p^2)$
differs by roughly a factor of 2 but the relation between 
$<g_3^2F^2>$ and $\sigma_{\rm fund}$ also changes and the 
result for $\tilde{\cal S}$ is close to that for $D^{(1)}$.}
As we can see the first term due to the spin interaction
dominates. The other terms related to minimal interaction have
opposite sign and go into the same direction as the
area law-induced mass square discussed in section 3.

Indeed, instead of the choice (\ref{A.28}) for the base
point $x_0$ one could choose  $x_0$ to be the center of mass 
introduced in eqs. (\ref{3.2}). The $\eta$-integrals together with 
the $\tau$-integrals in the first part of this appendix
then belong to area integrals filling the loop in the sense of the
nonabelian Stokes theorem. This is similar in spirit to ref. \cite{23}.
One could imagine a split of these
area integrals into segments related to the $\eta$-contributions
to (\ref{A.34}) and a main area integral approximated as in section 3
by a (modified) area law.

We mentioned already that\footnote{The minus sign is due
to the fact that the inverse dressed propagator is $G^{(0)-1}-\Sigma$.}
$\Delta m^2\equiv-\tilde{\cal S}(p^2=0)$ should be interpreted as the mass
shift due to the interaction with a stochastic background. This quantity
turns out to be IR-finite even in the case when the gauge bosons
are massless a priori, i.e., setting $m=0$ on the RHS of eq.
(\ref{A.34}) yields a finite function $\tilde{\cal S}(p^2)$ and in
particular a finite value for $\tilde{\cal S}(0)$. It is even possible
to calculate this value analytically. Returning to the
representation (\ref{A.30}), one obtains for arbitrary $m^2, d$
and $N_c$:
\bear\label{A.35}
&&-\Delta m^2=\int d^dz{\cal S}(z^2)=
\frac{\pi^{d/2}N_c}{d\Gamma(d/2)(N_c^2-1)}
<g^2F^2>\nonumber\\
&&\cdot\left\{\int^1_0d\eta_1d\eta_2\eta_1\eta_2\int^\infty_0dw\  
w^{d/2-1}wH'\left(\frac{w}{2}\right)D\left((\eta_1+\eta_2)^2
\frac{w}{4a^2}\right)\right.\nonumber\\
&&+\int^1_0d\eta_1d\eta_2(\eta_1\eta_2)^2\int^\infty_0dw\ w^{d/2-1}
(w/2a^2)H\left(\frac{w}{2}\right)D'\left((\eta_1
+\eta_2)^2\frac{w}{4a^2}\right)\nonumber\\
&&\left.+4\int^\infty_0dw\ w^{d/2-1}H\left(\frac{w}{2}\right)
D\left(\frac{w}{a^2}\right)\right\}\ear
Assuming now the propagator function $H$ to be of the massless
form (\ref{A.23a}), the integrals in (\ref{A.35}) can be
simplified considerably. In this case one may use the
identities $(d>2)$
\bear\label{A.36}
wH'\left(\frac{w}{2}\right)&=&-(d-2)H\left(\frac{w}{2}\right)
\nonumber\\
w^{d/2-1}H\left(\frac{w}{2}\right)&=&\frac{1}{4}\pi^{-d/2}
\Gamma(d/2-1)\ear
in order to bring (\ref{A.35}) to the following explicit form:
\bear\label{A.37}
-\Delta m^2&=&\frac{2N_c}{d(d-2)(N_c^2-1)}a^2<g^2F^2>\nonumber\\
&&\cdot\{1-(d-2)\theta_1-2\theta_2\}\int^\infty_0dw D(w)\ear
with the constants
\bear\label{A.38}
&&\theta_1\equiv\int^1_0d\eta_1\int^1_0d\eta_2\frac{\eta_1
\eta_2}{(\eta_1+\eta_2)^2}=\frac{2}{3}(1-\ln2)=0.2046...\nonumber\\
&&\theta_2\equiv\int^1_0d\eta_1\int^1_0d\eta_2\frac{(\eta_1
\eta_2)^2}{(\eta_1+\eta_2)^4}=\frac{1}{24}=0.0417...\ear
The remaining integral in (\ref{A.37}) slightly
depends on the precise shape of the model function $D$ we have
chosen. For the exponential $D(w)=e^{-w}$, corresponding
to eq. (\ref{2.10b}) the integral $\int^\infty_0dwD(w)$
equals unity, for instance. In any case it is a positive number
of order unity. 

Since in the physically interesting situation $d=
3$ the expression inside the curly brackets of (\ref{A.37}) is positive,
we are led to the important conclusion that $\Delta m^2$ is {\it negative}
(tachyonic). This shows that the interaction with the stochastic
background destabilizes the perturbative gauge field vacuum and
drives the system towards the formation of a condensate.

\section*{Appendix B}
\renewcommand{\theequation}{B.\arabic{equation}}
\setcounter{equation}{0}
In this appendix we explain the basic
mechanism of why stochastic background
fields lead to the formation of a condensate within
a simple scalar toy model.

Let $J(x_\mu)$ be a real scalar
Gaussian random variable, i.e. the average of an
arbitrary functional $F[J]$ is given by
\be\label{B.1}
<<F[J]>>=\int{\cal D} J\exp(-\frac{1}{2}\int J\Omega J)F[J]\ee
where
\be\label{B.2}
\int J\Omega J\equiv \int d^dxd^dy J(x)<x|\Omega|y>J(y)\ee
with some positive definite operator $\Omega$. Eq. (\ref{B.1})
implies that for any function $\phi(x)$
\be\label{B.3}
<<\exp\left(\pm\int d^dxJ(x)\phi(x)\right)>>
=\exp\left(+\frac{1}{2}\int\phi D\phi\right)\ee
where the operator $D\equiv\Omega^{-1}$ is positive
definite, too. We assume that $D$ (and $\Omega$)
is translational invariant with real matrix
elements $<x|D|y>=D(x-y)$. Then the only nonzero connected
correlation function reads
\be\label{B.4}
<<J(x)J(y)>>=D(x-y)\ee
This equation should be compared for the correlator
(\ref{1.3}) of the
stochastic vacuum model in QCD.

Let us consider the partition function $Z[J]$ of a scalar field
theory governed by the action $S[\phi]$ with an additional
linear coupling of the field $\phi(x)$ to a stochastic background
$J(x)$:
\be\label{B.5}
Z[J]=\int {\cal D}\phi\exp\left\{
-S[\phi]-\int d^dxJ(x)\phi(x)\right\}\ee
Its average is given by
\bear\label{B.6}
<<Z[J]>>&=&\int D\phi e^{-S[\phi]}<<e^{-\int J\phi}>>\nonumber\\
&=&\int {\cal D}\phi
\exp\left\{-S[\phi]+\frac{1}{2}\int \phi D\phi\right\}\ear
We see that a linear coupling of $\phi$ to a Gaussian random
field has the effect of changing the classical action according
to
\be\label{B.7}
S[\phi]\to S[\phi]-\frac{1}{2}\int d^dxd^dy\phi(x)D(x-y)\phi(y)\ee
It is important to observe that the new nonlocal term
in (\ref{B.7}) is negative for any $\phi(x)$. This means
in particular that it supplies a negative contribution to the
mass square of $\phi$. If $S$ is of the $Z_2$-symmetric
form $S[\phi]=\int d^dx\left\{\frac{1}{2}(\partial_\mu\phi)^2+\frac{1}{2}m^2\phi^2+
\frac{1}{4}\lambda \phi^4\right\}$, say, then the stochastic
background induces a mass shift
\be\label{B.8}
m^2\to m^2+\Delta m^2\ee
with
\be\label{B.9}
\Delta m^2=-\int d^dxD(x)=-\tilde D(p=0)<0\ee
where $\tilde D$ is the Fourier transform of $D$. If $m^2>0$
in absence of the $J\phi$ coupling, the theory has a $Z_2$-symmetric
vacuum with the minimum of the potential at $\phi=0$. If, however,
the coupling to the stochastic background is switched on
and if $|\Delta m^2|>m^2$, then $m^2+\Delta m^2$ becomes
negative. Hence the modified classical potential develops two
minima at $\phi\not=0$ and the $Z_2$-symmetry is spontaneously
broken.

In Yang-Mills theory a similar but technically less transparent
mechanism is at work. The Lagrangian for the gauge field
fluctuations $a_\mu(x)$ reads
$\frac{1}{2}a_\mu K_{\mu\nu}[A]a_\nu+O(a^3_\mu)$, and
clearly $a_\mu$ and $A_\mu$ are analogous to $\phi$ and $J$,
respectively. The coupling of $a_\mu$ to the
background $A_\mu$ is more complicated than the linear $J\phi$-term,
however. In particular, the analogue of the correlator (\ref{B.4})
is formulated in terms of $F_{\mu\nu}[A]$ rather than
$A_\mu$ itself. This problem can be overcome by
employing the Fock-Schwinger gauge which provides a simple
formula for $A_\mu$ in terms of $F_{\mu\nu}$, see eq. (\ref{2.5}). Both
the computation of the effective action in the main
body of the paper and the calculation of the mass operator
in Appendix A show that at least for $d=3$ the $\Delta m^2$ term
for $a_\mu$ which is induced by the stochastic background is negative,
see eq. (\ref{A.37}), for instance. For large values of $d$ we
find $\Delta m^2>0$, however.

\end{document}